\documentclass[final,3p,times]{elsarticle}

\usepackage{graphicx}
\usepackage{float}
\usepackage{tabularx}
\usepackage{adjustbox}
\usepackage{lipsum}
\usepackage{bbm}
\usepackage[dvipsnames]{xcolor,colortbl}

\usepackage{amsmath,mathrsfs,lineno,amssymb,,amsbsy}

\biboptions{compress}

\begin{document}

\begin{frontmatter}


\title{Theory of truncation resonances in continuum rod-based phononic crystals with generally asymmetric unit cells}

\author[address1,address2]{Hasan B. Al Ba'ba'a}
\author[address3,address4]{Carson L. Willey}
\author[address3,address4]{Vincent W. Chen}
\author[address3]{Abigail T. Juhl}
\author[address1]{Mostafa Nouh \corref{mycorrespondingauthor}}
\cortext[mycorrespondingauthor]{Corresponding author}
\ead{mnouh@buffalo.edu}

\address[address1]{
Dept. of Mechanical and Aerospace Engineering, University at Buffalo (SUNY), Buffalo, NY 14260, USA}
\address[address2]{
Department of Mechanical Engineering, Union College, Schenectady, NY 12308, USA}
\address[address3]{
Air Force Research Laboratory, Wright-Patterson AFB, OH 45433, USA}
\address[address4]{
UES, Inc., Dayton, OH 45432, USA}

\begin{abstract}
Phononic crystals exhibit Bragg bandgaps, frequency regions within which wave propagation is forbidden. In solid continua, bandgaps are the outcome of destructive interferences resulting from periodically alternating layers. Under certain conditions, natural frequencies emerge within these bandgaps in the form of high-amplitude localized vibrations near a structural boundary, referred to as \textit{truncation resonances}. In this paper, we investigate the vibrational spectrum of finite phononic crystals which take the form of a one-dimensional rod, and explain the factors that contribute to the origination of truncation resonances. By identifying a unit cell symmetry parameter, we define a family of finite phononic rods which share the same dispersion relation, yet distinct truncated forms. A transfer matrix method is utilized to derive closed-form expressions of the characteristic equations governing the natural frequencies of the finite system and decipher the truncation resonances emerging across different boundary conditions. The analysis establishes concrete connections between the localized vibrations associated with a truncation resonance, boundary conditions, and the overall configuration of the truncated chain as dictated by unit cell choice. The study provides tools to predict, tune, and selectively design truncation resonances, to meet the demands of various applications that require and uniquely benefit from such truncation resonances.
\end{abstract}

\end{frontmatter}

\section{Introduction}

Over the past few decades, vibration mitigation research has largely transitioned from traditional active control techniques and pursuit of materials with stronger damping properties to the notion of artificially engineered structures which possess a set of unique wave attenuation features that are otherwise elusive \cite{vasileiadis2021progress,dalela2022review}. Most notable among these are Phononic Crystals (PnCs), a class of periodic structures which come in continuous (e.g., rods \cite{AlBabaa2016a}, beams \cite{liu2012wave}, and plates \cite{Foehr2018Spiral-BasedInsulators}) or discrete (e.g., spring-mass chains \cite{Hussein2014} and lattices \cite{tanaka2000band}) forms. PnCs exhibit bandgaps, i.e., frequency regions of forbidden wave propagation, which arise from destructive interferences at interfaces with impedance mismatches. These mismatches are most commonly created by layering two or more materials in a spatially periodic manner with the smallest self-repeating block being referred to as the PnC's unit cell. Owing to their ability to manipulate incident vibroacoustic excitations, PnCs have been extensively used in a broad range of applications ranging from energy harvesting \cite{hu2021acoustic,lin2021piezoelectric} and attenuation of airborne sound \cite{bilal2018architected}, to nonreciprocal wave transmission \cite{attarzadeh2018non,attarzadeh2018elastic} and ultrasonic wave focusing \cite{danawe2020conformal}.

In their idealized form, PnCs are typically modeled as unbounded systems with an infinite number of unit cells. By applying the Bloch theorem, a dispersion diagram which consists of a set of curves relating the frequency of waves in the elastic medium to their wavenumber can be obtained \cite{Bloch1929}. These curves represent ``pass bands", or regions of permissible wave propagations, while empty spaces between them indicate bandgap regions where no real wavenumber solutions exist to satisfy a given frequency, culminating in an exponential attenuation of incident waves. Intuitively, natural frequencies (or resonances) of a vibrating PnC would therefore lie within the different pass bands of the dispersion relation. However, PnCs with a finite number of cells have been shown to exhibit natural frequencies inside bandgap regions, known as \textit{truncation resonances} -- named as such since they are a direct outcome of truncating the infinite medium and introducing boundary conditions \cite{AlBabaa2019DispersionCrystals}. While truncation resonances had been repeatedly observed in research involving PnCs, they were never explicitly investigated until 2011 \cite{davis2011analysis}. Mode shapes corresponding to truncation resonances are uniquely characterized by a profile of high vibrational amplitudes which are confined to the boundaries and decay away from it due to the bandgap effect. These profiles represent a set of ``edge" or ``surface" modes which have been instrumental in several emerging applications including phononic topological insulators \cite{susstrunk2015observation, Prodan2017DynamicalSystems, Rosa2019EdgeLattices}, flow control via phononic subsurfaces \cite{hussein2015flow, barnes2021initial}, and nano-particle mass sensing \cite{bonhomme2019love}.

Given the breadth and diversity of the number of engineering problems where localized attenuated modes play a central role, there is a strong motivation to study the dynamics of truncation resonances and comprehend their underlying features including onset conditions, patterns, distributions, and sensitivity to varying boundary conditions. Since truncation resonances are a function of finite PnCs, such goal can only be achieved by employing tools that go beyond the traditional Bloch-wave unit cell analysis, in order to capture the structural response of the finite system. Al~Ba'ba'a \textit{et al.} pioneered the use of continuous fractions \cite{hensley_contfrac} to devise a transfer-function-based model which depicts the frequency response of a finite spring-mass PnC \cite{al2017pole}, by utilizing a mathematical scheme that enables closed-form expressions for eigenvalues of perturbed tridiagonal matrices \cite{yueh2005eigenvalues,da2007eigenvalues}. The work set a framework to identify truncation resonance locations by recognizing natural frequencies which avert dispersion branches (i.e., fall outside the solutions of the analytical dispersion relation), and consequently, defined existence criteria for truncation resonances in free-free diatomic PnCs as a function of both mass and stiffness ratios as well as finite chain symmetry, as dictated by the truncation location. Additional studies of finite periodic systems shed light on additional discrepancies between the dispersion behavior of an infinite chain and the actual response of the final counterpart such as bandgaps which are undeveloped in terms of attenuation strength or frequency range \cite{al2017formation, al2018dispersion, sugino2016mechanism}. Bastawrous and Hussein extended the truncation resonance existence criteria to general boundary conditions as well as finite PnCs with end masses \cite{bastawrous2022closed}.

In this paper, we investigate the vibrational spectrum of finite continuum rod-based PnCs which consist of two alternating layers that repeat spatially to form a periodic composite-like structure. The primary goal is to provide the fundamental factors that contribute to the origination of truncation resonances. Contrary to their discrete (spring-mass) counterparts \cite{al2017pole, bastawrous2022closed}, closed-form equations of finite PnC rods cannot be obtained by using tridiagonal \textit{k}-Toeplitz matrices, and therefore warrant a different approach. In here, we will adopt the transfer matrix method (TMM) whose use in vibration characterization dates back to the 1950s, whether for flexural beams \cite{targoff1947associated} or elastic lattices \cite{maradudin1958vibrations,hori1957vibration,maradudin1958disordered}. Later on, and owing to its simplicity and efficiency as a predictive tool for finite vibrational structures, TMM has been widely adopted in the modeling of elastic periodic continua \cite{lin1969dynamics,mead1971vibration,xu2012low}. Using TMM, we will present an exhaustive analysis of the location, distribution, and associated wave propagation modes of the rod-based PnC's truncation resonances subject to all the possible boundary conditions. As eluded to in previous literature, the location at which an infinite periodic chain is truncated influences the overall symmetry of the resultant finite structure, and by extension, alters its truncation resonances, thus posing an additional variable to an already complex problem \cite{hvatov2015free, carneiro2021attenuation}. Furthermore, the final configuration of the truncated PnC often requires the use of a non-integer number of unit cells (For example, the analysis of a perfectly symmetric PnC of the form ``A-B-A-B-A" requires the use of 2.5 units cell of the form ``A-B"). The crux of the presented model lies in the choice of unit cell configuration, as depicted in Fig.~\ref{fig:PnC_UC}, which utilizes a single symmetry parameter $\delta$ to define the boundaries of the self-repeating portion. This streamlines the process of obtaining a family of finite PnC chains which share the same unit cell, yet distinct truncated forms, by boiling it down to a single design parameter made at the early stage of choosing the unit cell, and then repeating the chosen cell an integer number of times to build the finite structure. As such, the entire spectrum of possible truncated versions of a given PnC can be generated by sweeping the symmetry parameter between $-1$ and $1$.

Following a description of the parameters governing the design of a generally asymmetric unit cell, the TMM is used to derive expressions for the dispersion diagram as well as the natural frequencies, frequency response functions, and characteristic equations of the finite PnC for four possible boundary conditions, namely, free-free, fixed-fixed, fixed-free and free-fixed. By closely inspecting the first twelve bandgaps, we then explain the analogies and interplay between truncation resonances emerging across the four scenarios; making several references to the accompanying mode shapes and establishing connections between the localized vibrations and the corresponding boundary condition at that location. Finally, we track the truncation resonances and the aforementioned modes for each boundary condition as the unit cell configuration changes, enabling us to formalize the different patterns of truncation resonances that exist in a rod-based PnC of any truncated form. The framework presented here provides a pathway to predict and evaluate truncation resonances prior to fabrication, in addition to an ability to tune and selectively place their frequencies in a manner which caters to the different applications that require truncation resonances.

\section{Unit Cell Analysis}
\subsection{Asymmetric unit cell configuration \label{sec:unit_cell}}
\begin{figure*}[h!]
     \centering
\includegraphics[]{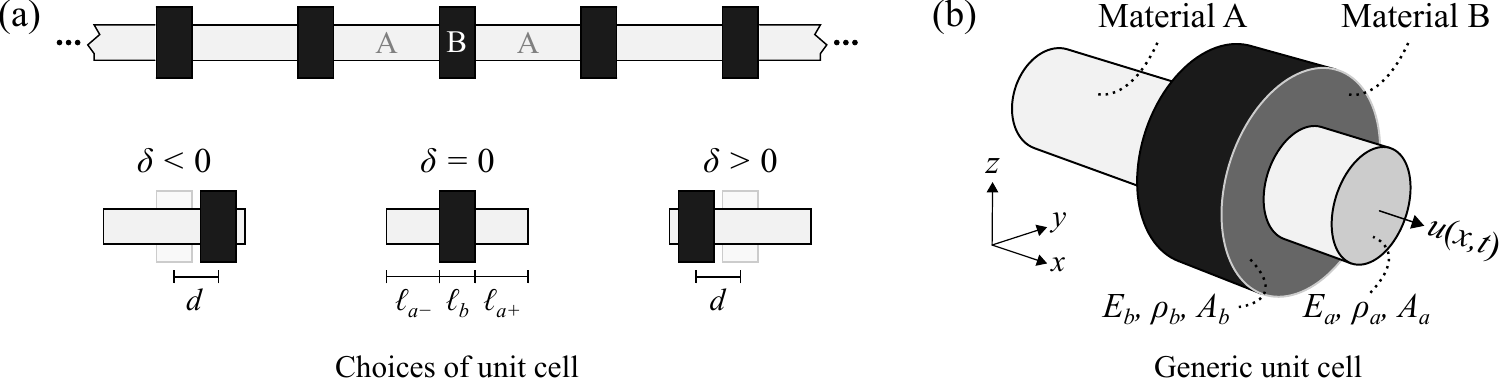}
     \caption{(a) Schematic diagram of a PnC rod of two alternating layers, A and B (top), and the different unit cell choices as a function of the symmetry parameter $\delta$ (bottom). (b) A three-dimensional  model of a generic unit cell indicating the material and geometric properties of the two constitutive layers, A and B. Longitudinal vibrations in the PnC rod are described by $u(x,t)$, where $t$ is time and $x$ is the axial direction.}
     \label{fig:PnC_UC}
\end{figure*}

Consider a one-dimensional PnC rod consisting of an infinite periodic arrangement of two segments, A and B, which are comprised of different materials and geometries, as shown in Fig.~\ref{fig:PnC_UC}. The rod's longitudinal vibrations at a given location $x$ at any time instant $t$ is given by the displacement function $u(x,t)$, as shown in Fig.~\ref{fig:PnC_UC}(b). Each of the two segments has an elastic modulus $E_{s}$, a density $\rho_{s}$, a cross-sectional area $A_{s}$, and a length $\ell_{s}$, where $s = a$ for segment A and $s=b$ for segment B. The self-repeating portion of the PnC rod, henceforth referred to as the unit cell, is chosen such that the total length of segment A is equal to $\ell_a = \ell_{a+} + \ell_{a-}$. The values of $\ell_{a+}$ and $\ell_{a-}$ dictate the level of symmetry of the unit cell as depicted in the lower panel of Fig.~\ref{fig:PnC_UC}(a). For any unit cell choice, a symmetry parameter $\delta$ is defined as
\begin{equation}
    \delta = \frac{2 d}{\ell_a},
\end{equation}
where $d = (\ell_{a+} - \ell_{a-})/2$ is the distance between the center of segment B in the chosen unit cell and that in a perfectly symmetric unit cell. As a result, the unit cell studied here is deemed asymmetric except when $\delta=0$. A choice of $\ell_{a+} > \ell_{a-}$ results in a positive $\delta$ while a choice of $\ell_{a+} < \ell_{a-}$ results in a negative $\delta$, with $\delta \in [-1,1]$. Using the definition of $d$, the lengths of $\ell_{a+}$ and $\ell_{a-}$ corresponding to a given symmetry parameter $\delta$ can be calculated as
\begin{equation}
    \ell_{a\pm} = \frac{\ell_a}{2} (1\pm \delta).
    \label{eq:la_pm}
\end{equation}

Equation~(\ref{eq:la_pm}) confirms the perfect unit cell symmetry condition, $\ell_{a+} = \ell_{a-}$, which takes place when $\delta = 0$. As the absolute value of $\delta$ increases, the asymmetry of the unit cell increases. At the limit of $|\delta| = 1$, the unit cell takes the form of a single continuous segment A of length $\ell_a$ which is either followed by segment B (when $\delta = -1$) or follows segment B (when $\delta = +1$). This limiting case of $|\delta| = 1$ is perhaps the most commonly used unit cell configuration in studies involving bi-layered PnCs (e.g., \cite{hussein2006dispersive,hussein2007dispersive,cheng2018complex}). However, the choice of $\delta$ and the effect it has on the shape of a finite PnC that consists of an integer number of unit cells (also known as the \textit{truncation} effect) is critical and at the heart of the present study. In this work, we will detail how the choice of $\delta$ dictates the presence of truncation resonances within the emerging bandgaps as well as the number and location of such resonances.

\subsection{Transfer matrix of a unit cell}

Throughout this paper, the transfer matrix method (TMM) is used as the primary analytical approach as it facilitates both the wave dispersion analysis based on a single unit cell, as well as the dynamics of a finite PnC rod made of a given number of cells which will be presented later. The transfer matrix $\mathbf{T}$ of a unit cell relates the components of the state vector, in this case the displacement $u$ and force $f$, at the end of one cell $i$ to the corresponding components at the end of an adjacent cell $i+1$, as follows \cite{Ruzzene2000}
\begin{equation}
    \begin{Bmatrix}
    u_{i+1} \\ f_{i+1}
    \end{Bmatrix} = \mathbf{T}
    \begin{Bmatrix}
    u_{i} \\ f_{i}
    \end{Bmatrix}.
    \label{eq:T_matrix}
\end{equation}

For the unit cell defined in Sec.~\ref{sec:unit_cell} with a symmetry parameter $\delta$, $\mathbf{T}$ is found by multiplying the transfer matrices for the three individual segments which form the unit cell, i.e.,
\begin{equation}
    \mathbf{T} = \mathbf{T}_{a_+} \mathbf{T}_b \mathbf{T}_{a_-},
    \label{eq:general_UC_T}
\end{equation}
where
\begin{equation}
    \mathbf{T}_{s} = 
    \begin{bmatrix}
    \cos(k_s \ell_{s} ) & \frac{1}{z_s \omega}\sin(k_s \ell_{s}) \\
    - z_s \omega\sin(k_s \ell_{s}) & \cos(k_s \ell_{s})
    \end{bmatrix}.
\end{equation}
with $s=a_+$, $b$, or $a_-$. Here, $z_{s} = A_{s} \sqrt{E_{s} \rho_{s}}$ is the characteristic impedance of each segment, and $k_{s} = \omega/c_{s}$ is the stand-alone wavenumber of materials A and B as a function of the angular frequency $\omega$ and the sonic speed of elastic medium $c_{s} = \sqrt{E_{s}/\rho_{s}}$. Note that $\delta = \pm 1$ renders one of the transfer matrices $\mathbf{T}_{a_\pm}$ an identity matrix, reducing Eq.~(\ref{eq:general_UC_T}) to two terms as expected. To streamline the analysis and reduce the number of parameters, we introduce $\omega_{s} = c_{s}/\ell_{s}$ as well as the two following non-dimensional quantities
\begin{subequations}
\begin{equation}
    \alpha =  \frac{\frac{1}{\omega_a}- \frac{1}{\omega_b}}{\frac{1}{\omega_a}+ \frac{1}{\omega_b}},
    \label{eq:alpha}
\end{equation}
\begin{equation}
    \beta = \frac{z_a - z_b}{z_a + z_b},
    \label{eq:beta}
\end{equation}
\end{subequations}
where $\alpha \in [-1,1]$ and $\beta \in [-1,1]$ are defined as the frequency and impedance contrasts, respectively. Using Eq.~(\ref{eq:beta}), the characteristic impedance of each material can be rewritten as
\begin{equation}
    z_{s} = z(1\pm \beta),
    \label{eq:z_beta}
\end{equation}
where $z = (z_a+z_b)/2$ is the average characteristic impedance of both materials. The plus ($+$) version of Eq.~(\ref{eq:z_beta}) denotes material A while the minus ($-$) version denotes material B. We also define a non-dimensional frequency $\Omega = \omega/\omega_0$, where $\omega_0$ is the harmonic mean of the frequencies $\omega_{a}$ and $\omega_{b}$ and is given by
\begin{equation}
    \omega_0 = \frac{2}{\frac{1}{\omega_a}+ \frac{1}{\omega_b}}.
\end{equation}
The harmonic mean $\omega_0$ can be combined with the definition of $\alpha$ to give
\begin{equation}
    \frac{1}{\omega_s} = \frac{1}{\omega_0} (1\pm\alpha),
    \label{eq:harmonic_alpha_combined}
\end{equation}
where, once again, the plus and minus versions of Eq.~(\ref{eq:harmonic_alpha_combined}) correspond to materials A and B, respectively. Consequently, the matrices in Eq.~(\ref{eq:general_UC_T}) can be expanded, resulting in the following unit cell transfer matrix
\begin{equation}
    \mathbf{T} = 
    \begin{bmatrix}
    t_{11} & t_{12} \\
    t_{21} & t_{22}
    \end{bmatrix},
    \label{eq:uc_TM}
\end{equation}
where
\begin{subequations}
\begin{align}
  t_{11} & = \cos(2\Omega \alpha) - \frac{2}{1-\beta^2} \sin \Big((1-\alpha)\Omega \Big)\Big[ \sin\Big((1+\alpha) \Omega\Big) - \beta \sin\Big(\delta(1+\alpha)\Omega \Big) \Big],
  \label{eq:t11} \\
  t_{12} & = \frac{\sin(2\Omega) + 2 \beta \sin\Big((1-\alpha)\Omega\Big) \cos\Big(\delta(1+\alpha)\Omega\Big) - \beta^2 \sin(2\Omega \alpha)}{z \omega (1-\beta)(1+\beta)^2},
    \label{eq:t12} \\
    t_{21} & = \frac{-z \omega}{1-\beta} \left [ \sin(2\Omega) - 2 \beta \sin\Big((1-\alpha)\Omega\Big) \cos\Big(\delta(1+\alpha)\Omega\Big) - \beta^2 \sin(2\Omega \alpha) \right], \text{and}
    \label{eq:t21} \\
    t_{22} & = \cos(2\Omega \alpha) - \frac{2}{1-\beta^2} \sin \Big((1-\alpha)\Omega\Big)\Big[ \sin\Big((1+\alpha) \Omega \Big) + \beta \sin\Big(\delta(1+\alpha)\Omega\Big)  \Big].
    \label{eq:t22}
\end{align}
\label{eq:UC_TM}
\end{subequations}

\subsection{Dispersion analysis}

Following the derivation of the transfer matrix $\mathbf{T}$, we obtain a closed-form expression for the PnC's dispersion relation. We start by computing the eigenvalues of the transfer matrix via $|\mathbf{T}-\lambda \mathbf{I}| = 0$. Using the fact that $|\mathbf{T}| = 1$, a unique property of transfer matrices, the characteristic equation can be found as
\begin{equation}
    \lambda^2 - \text{tr}(\mathbf{T}) \lambda + 1 = 0
    \label{eq:ch_eq_TM_UC},
\end{equation}
where $\text{tr}(\mathbf{T}) = t_{11} + t_{22}$ is the trace of matrix $\mathbf{T}$. The eigenvalues of $\mathbf{T}$ can be found by solving Eq.~(\ref{eq:ch_eq_TM_UC}), which yields the following solution pair
\begin{equation}
    \lambda_\pm = \frac{\text{tr}(\mathbf{T})}{2} \pm \sqrt{\left(\frac{\text{tr}(\mathbf{T})}{2} \right)^2 - 1}.
    \label{eq:lambda_roots}
\end{equation}
As such, it immediately follows that $\lambda_- + \lambda_+ = \text{tr}(\mathbf{T})$ from Eq.~(\ref{eq:lambda_roots}). Rewriting the eigenvalues as a function of a non-dimensional wavenumber $q$ in the form $\lambda_\pm = e^{\pm \mathbf{i} q}$, where $\mathbf{i}$ is the imaginary unit, gives
\begin{equation}
    \text{tr}(\mathbf{T}) = 2\cos(q),
    \label{eq:trace_cos(q)}
\end{equation}
from which the dispersion relation can be extracted as follows
\begin{equation}
    \cos(2 \Omega) - \beta^2 \cos (2\Omega \alpha) - (1-\beta^2) \cos(q) = 0.
    \label{eq:non-dim_disp_rel}
\end{equation}

It is evident from Eq.~(\ref{eq:non-dim_disp_rel}) that the dispersion relation is \textit{not} a function of the symmetry parameter $\delta$, which is a validating sign since the choice of the unit cell of an infinite PnC should not affect its dispersion diagram. Such choice, however, plays an instrumental role in the dynamics and eigenfrequency analysis of the finite PnC, as will be illustrated in Secs.~\ref{sec:Finite_PnC} and \ref{sec:TR}. The dispersion relation can be plotted analytically by calculating the non-dimensional wavenumber $q$ corresponding to any non-dimensional frequency $\Omega$ following a simple rearrangement of Eq.~(\ref{eq:non-dim_disp_rel})
\begin{equation}
   q = \cos^{-1} \left[ \frac{\cos(2 \Omega) - \beta^2 \cos (2\Omega \alpha)}{1-\beta^2} \right],
    \label{eq:non-dim_disp_driven_wave}
\end{equation}
which results in generally complex values of the wavenumber $q = q_\text{R} + \mathbf{i} q_\text{I}$. A bandgap emerges when $q_\text{I} \neq 0$ (or $|\lambda| \neq 1$) indicating spatial wave attenuation in the PnC rod. As can be inferred from Eq.~(\ref{eq:non-dim_disp_driven_wave}), a non-zero imaginary component of the wavenumber is guaranteed when the following inequality is satisfied
\begin{equation} \left| \frac{\cos(2 \Omega) - \beta^2 \cos (2\Omega \alpha)}{1-\beta^2} \right| > 1.
    \label{eq:BG_condition}
\end{equation}

\begin{figure*}[h!]
     \centering
\includegraphics[]{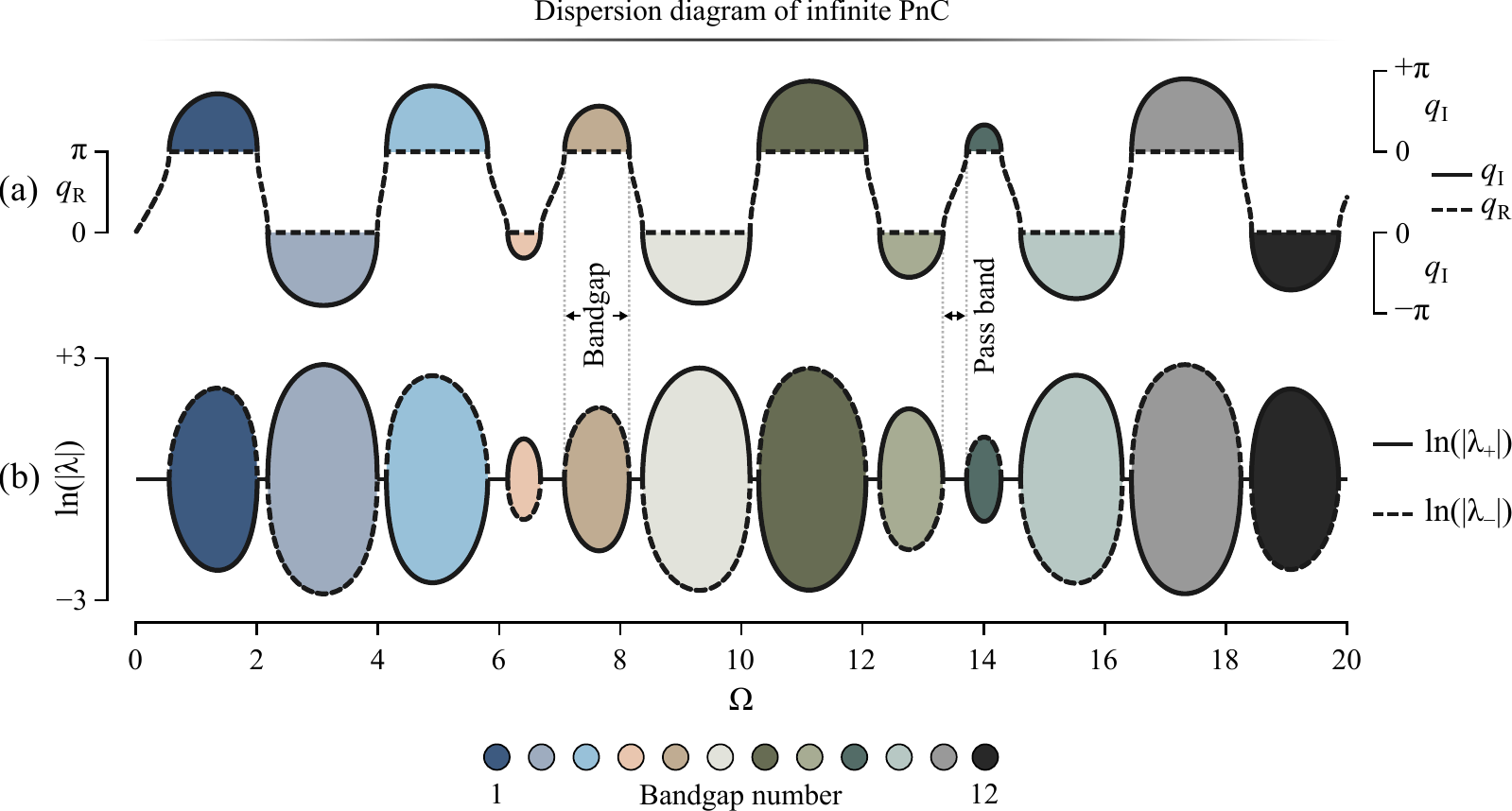}
     \caption{(a) Dispersion diagram of the PnC rod unit cell described in Table~\ref{table:material_properties}, and (b) the natural logarithm of the magnitude of the transfer matrix eigenvalues. The first twelve bandgaps are color coded to facilitate the interpretation of the forthcoming figures.}
     \label{fig:Dispersion}
\end{figure*}

Figure~\ref{fig:Dispersion}(a) shows the dispersion diagram for a PnC rod made of an ABS polymer (Material A) and Aluminum (Material B) corresponding to the parameters listed in Table~\ref{table:material_properties}. For these parameters, $\alpha$ and $\beta$ are found to be equal to $0.5379$ and $-0.8906$, respectively. The figure depicts the dispersion relation over the frequency range $\Omega \in [0,20]$, revealing the first twelve bandgaps of the PnC rod, which are shaded and color coded. In Fig.~\ref{fig:Dispersion}(b), the natural logarithm of the magnitude of the transfer matrix's eigenvalues is plotted, emphasizing that the eigenvalues are reciprocal of one another, i.e., $\lambda_+ \lambda_- = 1$ or $\text{ln}(|\lambda_+|)+\text{ln}(|\lambda_-|)= 0$. Frequency ranges corresponding to $\text{ln}(|\lambda|) \neq 0$ can be used as an alternative indicator of bandgap regions.

\begin{table*}[h!]
\caption{Material and geometric properties of segments A and B of the PnC unit cell.}
\centering
\begin{tabular}{l l c c c c}
\hline
& Material & Density & Young's Modulus & Area & Length \\
\hline
A & ABS & 1040 kg/$\text{m}^3$ & 2.4 GPa & 300 $\text{mm}^2$ & 40 mm \\
B & Aluminum & 2700 kg/$\text{m}^3$  & 69 GPa & 600 $\text{mm}^2$ & 40 mm \\
\hline
\end{tabular}
\label{table:material_properties}
\end{table*}

\section{Finite Structure Analysis}
\label{sec:Finite_PnC}
\subsection{Transfer matrix of a finite PnC rod}

The unit cell transfer matrix $\mathbf{T}$ derived in Eq.~(\ref{eq:uc_TM}) can be used to relate the displacement and forcing at the two ends of a finite PnC rod which consists of $n$ unit cells as follows
\begin{equation}
    \begin{Bmatrix}
    u_n \\ f_n
    \end{Bmatrix} = \mathbf{T}^n 
    \begin{Bmatrix}
    u_0 \\ f_0
    \end{Bmatrix},
    \label{eq:Tn_matrix}
\end{equation}
where $u_0$ and $f_0$ denote the displacement and force at $x=0$, $u_n$ and $f_n$ denote the displacement and force at $x = n \ell$, and $\ell = \ell_a + \ell_b$ represents the length of one unit cell. Alternatively, the matrix $\mathbf{T}^n$ can be found as a function of the unit cell's transfer matrix and the wavenumber $q$ using the following expression \cite{lin1969dynamics}
\begin{equation}
    \mathbf{T}^n = \frac{\cos(nq)}{2\cos(q)} \left(\mathbf{T} + \mathbf{T}^{-1} \right) + \frac{\sin(nq)}{2\sin(q)} \left(\mathbf{T} - \mathbf{T}^{-1} \right).
    \label{eq:Tn_1}
\end{equation}
Expanding the terms in Equation (\ref{eq:Tn_1}), it can be shown that
\begin{equation}
    \mathbf{T}^n = 
    \begin{bmatrix}
    \frac{\cos(nq)}{2\cos(q)} (t_{11}+t_{22}) + \frac{\sin(nq)}{2\sin(q)} (t_{11}-t_{22})  & \frac{\sin(nq)}{\sin(q)}t_{12} \\
    \frac{\sin(nq)}{\sin(q)}t_{21} & \frac{\cos(nq)}{2\cos(q)} (t_{11}+t_{22})
    - \frac{\sin(nq)}{2\sin(q)} (t_{11}-t_{22})
    \end{bmatrix}.
    \label{eq:Tn_2}
\end{equation}
Using Eq.~(\ref{eq:trace_cos(q)}) and the elements of the transfer matrix defined in Eq.~(\ref{eq:UC_TM}), the transfer matrix of a finite PnC with $n$ unit cells can be finally rewritten as
\begin{equation}
    \mathbf{T}^n = 
    \begin{bmatrix}
    t_{11n} & t_{12n} \\
    t_{21n} & t_{22n}
    \end{bmatrix},
    \label{eq:Tn_3}
\end{equation}
where
\begin{subequations}
\begin{align}
    t_{11n} & = \cos(nq) + \frac{2 \beta}{1-\beta^2} \frac{\sin(nq)}{\sin(q)} \Big[ \sin\Big((1-\alpha)\Omega\Big) \sin\Big(\delta(1+\alpha)\Omega\Big) \Big],
    \label{eq:t11n} \\
    t_{12n} & = \frac{1}{z \omega (1-\beta)(1+\beta)^2} \frac{\sin(nq)}{\sin(q)} \left(\sin(2\Omega) + 2 \beta \sin\Big((1-\alpha)\Omega\Big) \cos\Big(\delta(1+\alpha)\Omega\Big) - \beta^2 \sin(2\Omega \alpha) \right), \label{eq:t12n} \\
    t_{21n} & = - \frac{z \omega }{1-\beta} \frac{\sin(nq)}{\sin(q)} \left( \sin(2\Omega) - 2 \beta \sin\Big((1-\alpha)\Omega\Big) \cos\Big(\delta(1+\alpha)\Omega\Big) - \beta^2 \sin(2\Omega \alpha) \right), \text{and} \\
    t_{22n} & = \cos(nq) - \frac{2 \beta}{1-\beta^2} \frac{\sin(nq)}{\sin(q)} \Big[ \sin\Big((1-\alpha)\Omega\Big) \sin\Big(\delta(1+\alpha)\Omega\Big) \Big].
    \label{eq:t22n}
\end{align}
\end{subequations}

\subsection{End-to-end frequency response functions for different boundary conditions}
\label{sec:FRF}

By parsing the different elements of the matrix $\mathbf{T}^n$, a set of frequency response functions (FRFs) which describe the ratio between the displacements at both ends of the finite PnC rod (e.g., $u_n/u_0$ or $u_0/u_n$), or the displacement at one end to the force applied at the other end (e.g., $u_n/f_0$) can be obtained for commonly encountered boundary conditions. First, Eq.~(\ref{eq:Tn_matrix}) is expanded into the following two equations
\begin{subequations}
\begin{equation}
    u_n = t_{11n} u_0 + t_{12n} f_0,
    \label{eq:Tn_expanded_1}
\end{equation}
\begin{equation}
    f_n = t_{21n} u_0 + t_{22n} f_0.
    \label{eq:Tn_expanded_2}
\end{equation}
\end{subequations}

For a free-free rod, a force $f_0$ can be prescribed at $x=0$ with the displacement measured at $x=n\ell$. This implies that $f_n = 0$ which results in $u_0 = - t_{22n}/t_{21n} f_0$ as can be obtained from Eq.~(\ref{eq:Tn_expanded_2}). Substituting back into Eq.~(\ref{eq:Tn_expanded_1}) and taking advantage of the fact that the determinant of $\mathbf{T}_n$ is unity, the following FRF can be computed
\begin{equation}
    \frac{u_n}{f_0} =- \frac{1}{t_{21n}}.
    \label{eq:TF_FreeFree}
\end{equation}

For fixed-free and free-fixed rods, the FRF can be obtained via a base excitation of the fixed end and a displacement measurement at the free one. In the fixed-free case, we prescribe a base displacement $u_0$ and set $f_n = 0$ for the free end at $x = n\ell$. Following which, we solve for the force $f_0 = -t_{21n}/t_{22n} u_0$ from Eq.~(\ref{eq:Tn_expanded_2}) and plug that into Eq.~(\ref{eq:Tn_expanded_1}) to get
\begin{equation}
\frac{u_n}{u_0} = \frac{1}{t_{22n}}.
\end{equation}
Analogously, the FRF for the free-fixed rod can be obtained by following a similar procedure to that of the fixed-free rod but switching the base excitation and measurement locations, thus resulting in
\begin{equation}
\frac{u_0}{u_n} = \frac{1}{t_{11n}}.
\end{equation}

\begin{figure*}[h!]
     \centering
\includegraphics[]{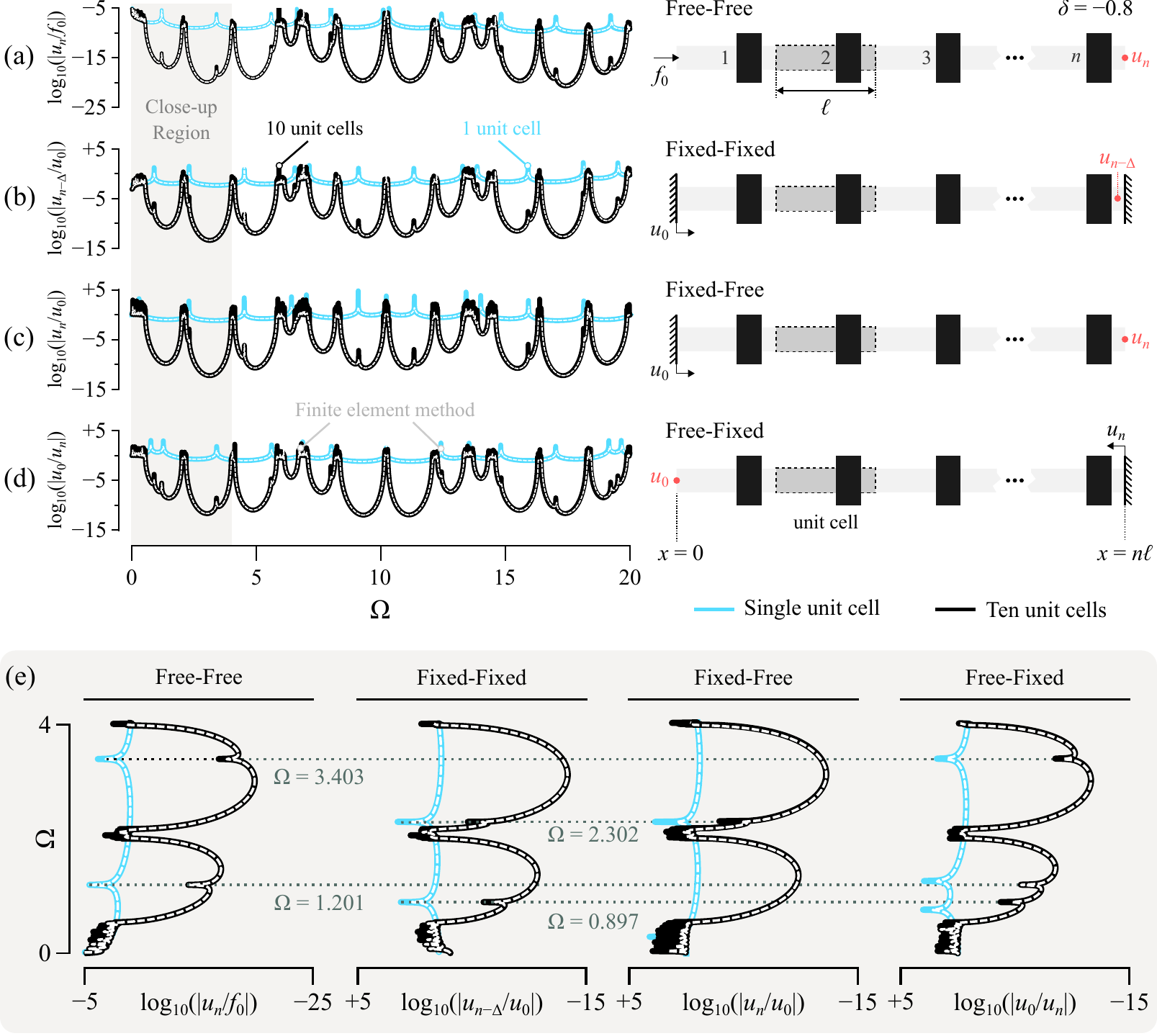}
     \caption{Frequency response functions (FRFs) of finite PnC rods comprised of a single unit cell ($n=1$; solid blue) and 10 unit cells ($n=10$; solid black), both corresponding to a symmetry parameter of $\delta = -0.8$, analytically obtained via TMM. FEM results for both cases are superimposed on all the results (white dashed curves). The FRFs are presented for four boundary conditions as follows: (a) Free-free, (b) Fixed-fixed, (c) Fixed-free, and (d) Free-fixed. The rightmost column panel provides illustrative schematics of the different excitations and boundary conditions. (e) A close-up of the FRFs shown in (a-d) for the shaded region $\Omega \in [0,4]$, showing natural frequencies appearing within the first two bandgaps.}
     \label{fig:FRF}
\end{figure*}

A slightly different treatment is required to attain the FRF of the fourth, and final, boundary condition, namely the fixed-fixed rod. Since both ends of the rod are fixed and a direct measurement of the displacement at either end is not possible, the FRF will be obtained as a ratio between the displacement at a point very close to one of the fixed ends, $u_{n-\Delta}$, which is measured at $x = n\ell-\Delta$, and a base excitation $u_0$ at the opposite end. Here, $\Delta$ is a distance parameter. As a result, the state vector at $x = n\ell-\Delta$ can be found by a transfer matrix inverse as follows
\begin{equation}
    \begin{Bmatrix}
    u_{n-\Delta} \\ f_{n-\Delta}
    \end{Bmatrix}
    = 
    \mathbf{T}_\Delta^{-1}
    \begin{Bmatrix}
    u_{n} \\ f_{n}
    \end{Bmatrix},
    \label{eq:T_Delta}
\end{equation}
where
\begin{equation}
    \mathbf{T}_\Delta^{-1} = 
    \begin{bmatrix}
    \cos\left(\frac{\omega}{c_a} \Delta \right) & -\frac{1}{\omega z_a} \sin \left(\frac{\omega}{c_a} \Delta \right) \\
    \omega z_a \sin \left(\frac{\omega}{c_a} \Delta \right)  & \cos\left(\frac{\omega}{c_a} \Delta \right) \\ 
    \end{bmatrix}.
\end{equation}
Note that consistent with our generalized unit cell definition in Sec.~\ref{sec:unit_cell} and Fig.~\ref{fig:PnC_UC}, both ends of the finite structure are made out of material A for all unit cell choices (except for the one unit cell design corresponding to $\delta = -1$ which ends with material B), thus justifying the use of material A properties in the $\mathbf{T}_\Delta^{-1}$ expression. Assuming a very small $\Delta$, $\mathbf{T}_\Delta^{-1}$ can be linearized about $\Delta=0$ as follows
\begin{equation}
    \mathbf{T}_{\Delta,0}^{-1} = 
    \begin{bmatrix}
    1 & -\frac{1}{z_a c_a} \Delta \\
     \frac{\omega^2 z_a}{c_a} \Delta & 1 \\ 
    \end{bmatrix}.
\end{equation}
Consequently, the state vector right before the fixed end (i.e., at $x = n\ell - \Delta$) can alternatively be computed using Eq.~(\ref{eq:T_Delta}), albeit with $\mathbf{T}^{-1}_{\Delta,0}$ instead of $\mathbf{T}^{-1}_{\Delta}$. Setting $u_n =0$ at the fixed end and using Eqs.~(\ref{eq:Tn_expanded_1}) and~(\ref{eq:Tn_expanded_2}) to derive an expression for $f_n$ as a function of $u_0$, we obtain $f_n = -u_0/t_{12n}$, and thus the FRF for this boundary condition can be shown to be
\begin{equation}
    \frac{u_{n-\Delta}}{u_0} = \frac{\Delta(1+\alpha)}{\ell_a \omega_0 z (1+\beta)} \frac{1}{t_{12n}}.
    \label{eq:fixed_fixed_FRF}
\end{equation}
Note that setting $\Delta = 0$ in Eq.~(\ref{eq:fixed_fixed_FRF}) recovers the boundary condition $u_n = 0$. Using the same properties listed in Table~\ref{table:material_properties} and $\delta = -0.8$, the FRFs for all four boundary conditions for a finite PnC rod made out of a single unit cell ($n=1$; blue curves) and ten unit cells ($n=10$; black curves) are plotted in Fig.~\ref{fig:FRF}. Furthermore, to validate the analytical expressions obtained earlier, the same FRFs are also reproduced via a finite element model (FEM) of the finite structures, coded via MATLAB using two-node rod elements \cite{petyt2010introduction}. For the FEM results, the number of finite elements per unit cell is $n_e = 250$, and $\Delta = \ell/n_e$ is used in the fixed-fixed case. All the FEM plots are superimposed on the analytical plots for comparison (white dashed lines), and an excellent agreement between the analytical and FEM data can be seen.

Two important observations are made regarding Bragg bandgaps exhibited by the finite PnC rod. The first is that such bandgaps are recognized in FRFs as regions of significant u-shaped drops in amplitude, which are more pronounced in structures comprising a larger number of unit cells (e.g., easier to detect in the $n=10$ FRFs than the $n=1$ FRFs in Fig.~\ref{fig:FRF}). The formation mechanism of such bandgaps in finite PnCs and the underlying reasons behind the extent of amplitude drop and the number of unit cells are thoroughly detailed in Ref.~\cite{al2017pole}. The second observation pertains to the appearance of resonant peaks inside some bandgaps in the $n=10$ case, often referred to as truncation \cite{AlBabaa2019DispersionCrystals, al2017pole} or bandgap resonances \cite{bastawrous2022closed}. Interestingly, for both the free-free and fixed-fixed cases, we observe that these truncation resonances perfectly coincide with the natural frequencies of the single unit cell which is not the case for the two other boundary conditions. Even more intriguingly, we note that while the truncation resonances in the fixed-free and free-fixed cases do not necessarily align with the natural frequencies of the single unit cell in their respective cases, each of these truncation resonances perfectly coincides with a single cell natural frequency from either the free-free or fixed-fixed cases, as highlighted in the close-up shown in Fig.~\ref{fig:FRF}(e). The locations and unique features of these truncation resonances will be confirmed and analytically proven in the subsequent sections by closely examining the characteristic equations of the finite PnCs subject to different boundary conditions.

\section{Truncation Resonances}
\label{sec:TR}
\subsection{Natural frequencies for different boundary conditions \label{sec:Nat_Freq_BCs}}

To find the natural frequencies of the finite PnC rod, we set the force or displacement to zero for fixed or free ends, respectively, and use either Eq.~(\ref{eq:Tn_expanded_1}) or (\ref{eq:Tn_expanded_2}) to find the characteristic equation. For example, $f_0 = f_n = 0$ in a free-free rod. Substituting these conditions into Eq.~(\ref{eq:Tn_expanded_2}), a non-trivial solution is only possible if $t_{21n} = 0$, which constitutes the characteristic equation for the free-free case. The solutions of the characteristic equation represent the finite structure's natural frequencies. It should be noted that these solutions also represent the roots of the denominator of the transfer function in Eq.~(\ref{eq:TF_FreeFree}), which are commonly known as the system's \textit{poles} in controls and linear systems theory \cite{curtain2009transfer}. A similar process can be adopted for the rest of the boundary conditions to obtain the remaining characteristic equations, which are summarized in Table~\ref{table_FRFs_ChEqs} for ease of reference.

\begin{table*}[h!]
\caption{Characteristic equations and FRFs of commonly encountered boundary conditions of the finite PnC rod.}
\centering
\begin{tabular}{l l l}
\hline
Boundaries & Characteristic Equation & Frequency Response Function \\
\hline
Free-Free & $t_{21n} = 0$ & $\frac{u_n}{f_0} =- \frac{1}{t_{21n}}$\\[0.5em]
Fixed-Fixed & $t_{12n} = 0$ & $\frac{u_{n-\Delta}}{u_0} = \frac{\Delta(1+\alpha)}{\ell_a \omega_0 z (1+\beta)} \frac{1}{t_{12n}}$\\[0.5em]
Fixed-Free & $t_{22n} = 0$ & $\frac{u_0}{u_n} = \frac{1}{t_{22n}}$\\[0.5em]
Free-Fixed & $t_{11n} = 0$ & $\frac{u_0}{u_n} = \frac{1}{t_{11n}}$ \\[0.75em]
\hline 
\end{tabular}
\label{table_FRFs_ChEqs}
\end{table*}

\subsubsection{Free-free and fixed-fixed cases}
We begin with the free-free case with the characteristic equation $t_{21n} = 0$, which reads
\begin{equation}
    \Omega \frac{\sin(nq)}{\sin(q)} \left( \sin(2\Omega) - 2 \beta \sin\Big((1-\alpha)\Omega\Big) \cos\Big(\delta(1+\alpha)\Omega\Big) - \beta^2 \sin(2\Omega \alpha)  \right) = 0.
    \label{eq:free-free_Ch_Eq}
\end{equation}
From Eq.~(\ref{eq:free-free_Ch_Eq}), it can be inferred that the solution $\Omega = 0$, which describes a non-oscillatory rigid body mode, satisfies the characteristic equation, as expected in a free-free (unconstrained) structure. Additionally, the roots obtained from $\frac{\sin(nq)}{\sin(q)} = 0$ provide closed-form wavenumber solutions of the following form
\begin{equation}
    q = \frac{r \pi}{n},
    \label{eq:q_solutions_free-free}
\end{equation}
for all integer values of $r$ except zero and multiples of $n$, which yield indeterminate expressions. These values can be then substituted into the dispersion relation in Eq.~(\ref{eq:non-dim_disp_rel}), to get
\begin{equation}
    \cos(2 \Omega) - \beta^2 \cos (2\Omega \alpha) - (1-\beta^2) \cos\left(\frac{r\pi}{n}\right) = 0.
    \label{eq:natfreq_freefree_fixedfixed}
\end{equation}
Given the periodicity of the dispersion relation, the range $q \in [0,\pi]$ defining the irreducible Brillouin zone suffices for the analysis. As a result, the integer $r$ can be limited to $r = 1,2, \dots, n-1$. Equation~(\ref{eq:natfreq_freefree_fixedfixed}) can then be solved numerically to compute the subset of the PnC's natural frequencies which satisfy the dispersion relation, i.e., the natural frequencies which lie within pass band regions and correspond to the purely real values of $q$ in Eq.~(\ref{eq:q_solutions_free-free}). Each value of the integer $r$ will result in an infinite number of natural frequency solutions, one per dispersion branch, regardless of the unit cell symmetry parameter $\delta$. The fact that there exists an infinite number of natural frequencies is a hallmark feature of finite continuum structures, contrary to finite spring-mass lattices which comprise a finite number of degrees of freedom and hence a finite number of resonant vibrational modes \cite{AlBabaa2019DispersionCrystals, al2017pole}. Finally, the presence of resonant peaks outside of pass bands (and inside bandgaps) in Fig.~\ref{fig:FRF}(a) suggests that another set of natural frequencies must emanate from the roots of the third factor in Eq.~(\ref{eq:free-free_Ch_Eq}), representing the truncation resonances which correspond to the free-free boundary condition. These are given by the solutions of
\begin{equation} \sin(2\Omega) - 2 \beta \sin\Big((1-\alpha)\Omega\Big) \cos\Big(\delta(1+\alpha)\Omega\Big) - \beta^2 \sin(2\Omega \alpha) = 0.
    \label{eq:possible_TP}
\end{equation}

Here, we confirm the observation made earlier in Fig.~\ref{fig:FRF}(a) that the truncation resonances of a free-free PnC rod match the natural frequencies of a single unit cell with identical boundary conditions. This feature is easily provable by using Eq.~(\ref{eq:free-free_Ch_Eq}) and setting the number of cells $n$ equal to one, and is inline with earlier results reported for mono-coupled systems~\cite{mead1975wave}. It also serves as both a valuable and efficient tool in applications requiring a priori placement of truncation resonances, rather than arbitrary trial and error. It is, however, important to reiterate that the roots of Eq.~(\ref{eq:possible_TP}) may not always lie inside a bandgap and could, in some cases, end up satisfying a dispersion branch. Whether or not a given solution lie within a bandgap can always be checked via the inequality provided in Eq.~(\ref{eq:BG_condition}). As such, it is prudent to refer to the solutions of Eq.~(\ref{eq:possible_TP}) as \textit{potential} truncation resonances. Simply stated, not every solution of Eq.~(\ref{eq:possible_TP}) is necessarily a truncation resonance but any truncation resonance is a solution of Eq.~(\ref{eq:possible_TP}).

\begin{figure*}[ht]
     \centering
\includegraphics[width=\textwidth]{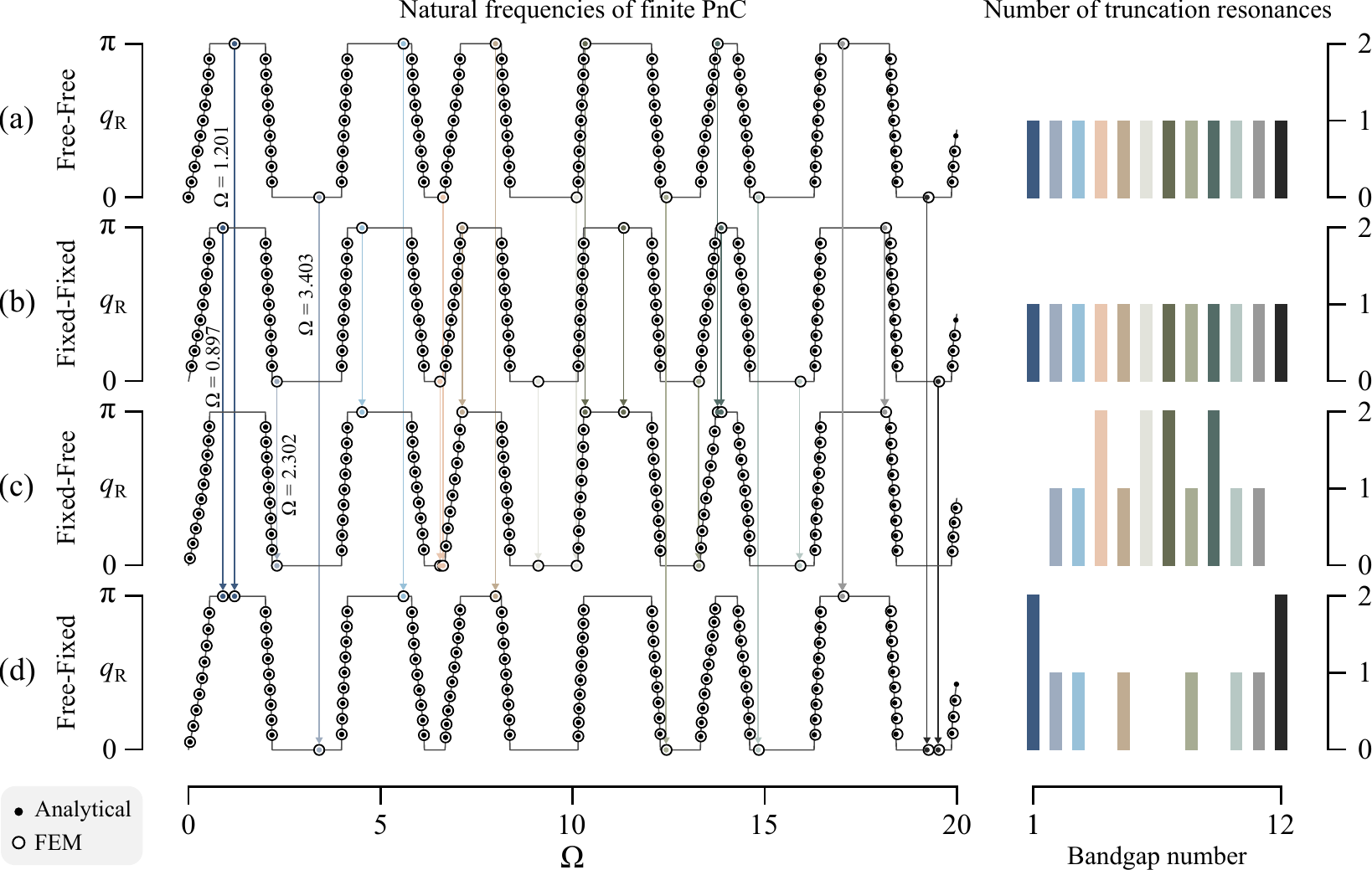}
     \caption{Analytical (circles) and FEM-obtained (dots) natural frequencies for finite PnC rod subject to the following boundary conditions: (a) Free-free, (b) Fixed-fixed, (c) Fixed-free, and (d) Free-fixed. The natural frequencies are superimposed on the dispersion relation for reference. The rightmost column shows the number of natural frequencies that appear in each bandgap, which are color coded to facilitate figure interpretation and correspond to the legend in Fig.~\ref{fig:Dispersion}.}
     \label{fig:Dispersion_natfreq}
\end{figure*}

Next, we move to the fixed-fixed boundary condition where $t_{12n} = 0$ represents the characteristic equation of the finite system. From Eq.~(\ref{eq:t12n}), it can be inferred that all the non-zero pass band natural frequencies of the fixed-fixed PnC rod are identical to those of the free-free one, and can be similarly obtained using Eq.~(\ref{eq:natfreq_freefree_fixedfixed}). However, the potential truncation resonances in this case need to be obtained from the following equation
\begin{equation} \sin(2\Omega) + 2 \beta \sin\Big((1-\alpha)\Omega\Big) \cos\Big(\delta(1+\alpha)\Omega\Big) - \beta^2 \sin(2\Omega \alpha) = 0.
    \label{eq:possible_TP2}
\end{equation}
It is important to note that the only distinguishing feature between Eqs.~(\ref{eq:possible_TP}) and (\ref{eq:possible_TP2}) is the sign of the second term, meaning that if the sign of $\beta$ is flipped (i.e., impedance values are swapped), the truncation resonances of a fixed-fixed rod will become those of a free-free one, and vice versa. Equally important is the fact that the sign of $\delta$ has no effect on the truncation resonances here since it is a part of the argument of the cosine function in both equations.

Using the same parameters as those used to generate Fig.~\ref{fig:FRF}, all the natural frequencies of the free-free and fixed-fixed PnC rods up to $\Omega=20$ are computed analytically using the equations derived here, and projected on the dispersion diagram, as shown in Figs.~\ref{fig:Dispersion_natfreq}(a) and \ref{fig:Dispersion_natfreq}(b), respectively. Additionally, and for verification, the natural frequencies are also computed via the same FEM model described in Sec.~\ref{sec:FRF}, and superimposed on the figures for comparison; generally showing excellent agreement with the analytical values with minor deviations at higher frequencies consistent with FEM resolution. The right column of Figs.~\ref{fig:Dispersion_natfreq}(a) and (b) shows a count of the number of truncation resonances that exist in each bandgap, revealing a constant total of one truncation resonance per bandgap in both cases. For convenience, truncation resonance features (e.g., value and count) follow the same color code used in Fig.~\ref{fig:Dispersion}, allowing each truncation resonance to be easily traced to its corresponding bandgap. 

Finally, we note that while the analytical derivations detailed here are established for a PnC rod comprising two different materials, the findings are generalizable for any free-free and fixed-fixed PnC regardless of the number of material layers. A simple proof can be obtained by inspecting the form of the matrix $\mathbf{T}^n$ and by comparing Eqs.~(\ref{eq:Tn_2}) and (\ref{eq:Tn_3}). It can be seen that $t_{21n}$ is found from the multiplication of the terms $\frac{\sin(nq)}{\sin(q)}$ and $t_{21}$. Setting $t_{21n} = 0$ also mandates that $t_{21} = 0$, suggesting that the natural frequencies of a free-free PnC with $n=1$ represent a subset of a free-free PnC natural frequencies with any $n>1$. By replacing $t_{21n}$ with $t_{12n}$, the same can be said about the fixed-fixed PnC rod. This conclusion remains true regardless of the number of material layers constituting the unit cell, a fact that was also proven for a general class of free-free polyatomic PnC lattices~\cite{AlBabaa2019DispersionCrystals}.

\begin{figure*}[h!]
     \centering
\includegraphics[]{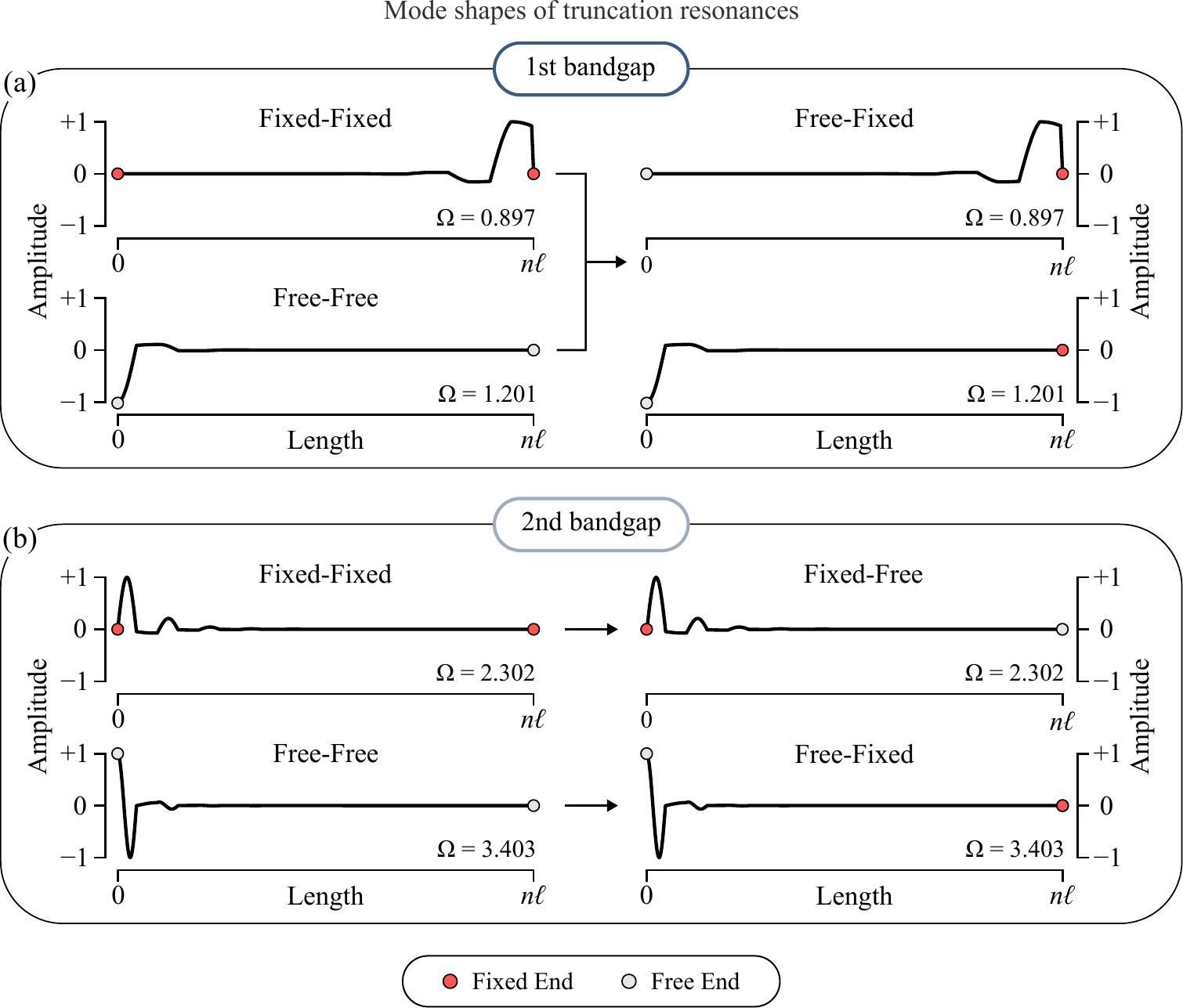}
     \caption{Modes shapes corresponding to truncation resonances revealing edge modes which attenuate away from the localized vibrations, provided as follows: (a) First bandgap for fixed-fixed, free-free, and free-fixed boundary conditions, and (b) second bandgap for fixed-fixed, free-free, fixed-free, and free-fixed boundary conditions.}
     \label{fig:mode_shapes}
\end{figure*}

\subsubsection{Fixed-Free and Free-Fixed cases \label{sec:Fixed_Free_trunc}}

As indicated in Table~\ref{table_FRFs_ChEqs}, the characteristic equations of the free-fixed and fixed-free finite PnC rods are $t_{11n}=0$ and $t_{22n}=0$, respectively, and their explicit forms are found by setting Eqs.~(\ref{eq:t11n}) and (\ref{eq:t22n}) equal to zero. By closely examining these two characteristic equations, it can be also observed that they can be obtained from one another by either changing the sign of the contrast parameter $\beta$ or the symmetry parameter $\delta$. Unlike the free-free and fixed-fixed cases, an analytical expression for the wavenumber $q$ is generally not available for general system parameters. As such, the natural frequencies can be found by plugging in Eq.~(\ref{eq:non-dim_disp_driven_wave}) into $t_{11n}=0$ or $t_{22n}=0$, and numerically finding $\Omega$ solutions that satisfy the resulting equation. This approach yields all the natural frequencies, whether they represent pass bands or truncation resonances. Following which, the natural frequencies $\Omega$ can be used to determine the corresponding wavenumber from Eq.~(\ref{eq:non-dim_disp_driven_wave}). Figure~\ref{fig:Dispersion_natfreq}(c,d) shows all the natural frequencies of the fixed-free and fixed-free PnC rods up to $\Omega=20$. By comparing the results with those obtained earlier for the free-free and fixed-fixed cases in Fig.~\ref{fig:Dispersion_natfreq}(a,b), we observe that each truncation resonance that appears in either the fixed-free or free-fixed cases must already exist in the free-free or fixed-fixed ones. This feature is graphically emphasized in Fig.~\ref{fig:Dispersion_natfreq} via a series of arrows which trace each truncation resonance in the figure's last two rows to its counterpart in the first two rows. 

To illustrate the underlying dynamics of truncation resonances associated with these two boundary conditions, we consider the first bandgap in the free-fixed case. As depicted in Fig.~\ref{fig:Dispersion_natfreq}(d), two truncation resonances appear in this bandgap. By tracing the arrows, the first one at $\Omega=0.897$ corresponds to a truncation resonance from the fixed-fixed case, while the second one at $\Omega=1.201$ corresponds to a truncation resonance from the free-free case. This observation can be further explained by considering the eigenvectors associated with these two truncation resonances, or their mode shapes, in Fig.~\ref{fig:mode_shapes}(a). Since these are resonant modes that appear within a bandgap, i.e., a region of spatial wave attenuation, mode shapes of truncation resonances typically show vibrations which are localized or confined to one end of the structure \cite{AlBabaa2019DispersionCrystals}. At $\Omega = 0.897$, the mode shapes of the fixed-fixed and free-fixed systems resemble each other and show localized vibrations near the rightmost end of the rod, which is fixed in both cases -- explaining the matching truncation resonances. Similarly, at $\Omega = 1.201$, the mode shapes of the free-free and free-fixed systems resemble each other and show localized vibrations near the leftmost end of the rod, which is free in both cases. Additional examples of matching truncation resonances which lie in the second bandgap are provided in Fig.~\ref{fig:mode_shapes}(b). One additional observation from Fig.~\ref{fig:mode_shapes}(a) is that the displacement profiles at $\Omega = 0.897$ exhibit a slower spatial decay of vibrations away from the localized edge when compared to their counterparts at $\Omega = 1.201$. This is attributed to the strength of bandgap attenuation within the first bandgap at each of these two frequencies, as dictated by the imaginary component of the wavenumber $q_\text{I}$, and shown in Fig.~\ref{fig:Dispersion}(a). The same observation can be made for truncation resonance mode shapes falling within the second bandgap, shown in Fig.~\ref{fig:mode_shapes}(b).

Interestingly, an analytical proof for the observations above can be derived by considering a semi-infinite PnC, similar to finding surface modes in elastic lattices \cite{djafari1983sagittal} or photonic crystals \cite{shukla2015properties}. We propose the following proof using a combination of the transfer matrix of the finite PnC rod and the solution of the displacement and internal forces within an elastic medium. Starting with the latter, the displacement and internal force at an arbitrary point $x$ along a rod with longitudinal vibrations can be found using a combination of forward- and backward-propagating waves as follows
\begin{subequations}
\begin{equation}
    \hat{u}(x) = a_1 \text{e}^{-\textbf{i}k_s x} + a_2 \text{e}^{\textbf{i}k_s x},
    \label{eq:u_equ}
\end{equation}
\begin{equation}
    \hat{f}(x) = -\textbf{i} E_s A_s k_s \left(a_1 \text{e}^{-\textbf{i}k_s x} -  a_2 \text{e}^{\textbf{i}k_s x} \right).
    \label{eq:f_equ}
\end{equation}
\end{subequations}
At $x = 0$, we have $u_0 = \hat{u}(0)$ and $f_0 = \hat{f}(0)$ and using Eqs.~(\ref{eq:u_equ}) and (\ref{eq:f_equ}), a compact matrix form of the two equations can be written as
\begin{equation}
    \begin{Bmatrix}
    u_0 \\ f_0
    \end{Bmatrix} = 
    \underbrace{
    \begin{bmatrix}
    1 & 1 \\
    -\textbf{i}E_s A_s k_s & \textbf{i}E_s A_s k_s 
    \end{bmatrix}}_{\mathbf{U}_s}
    \begin{Bmatrix}
    a_1 \\ a_2
    \end{Bmatrix}.
    \label{eq:coeff_matrix0}
\end{equation}
Note that zeroing out the displacement (or force) in Eq.~(\ref{eq:coeff_matrix0}) implies that $a_1 = -a_2$ (or $a_1 = a_2$). Now, let us revisit Eq.~(\ref{eq:Tn_matrix}) which relates the state vectors at $x = 0$ and $x = n\ell$. If we consider a semi-infinite medium, we can relate these two vectors using the eigenvalues of the transfer matrix as
\begin{equation}
    \begin{Bmatrix}
    u_n \\ f_n
    \end{Bmatrix} =
    \lambda^n
    \begin{Bmatrix}
    u_0 \\ f_0
    \end{Bmatrix}.
    \label{eq:Tn_semi-infinite}
\end{equation}
Using Eqs.~(\ref{eq:Tn_matrix}),~(\ref{eq:coeff_matrix0}) and~(\ref{eq:Tn_semi-infinite}), we arrive at an eigenvalue problem of the form
\begin{equation}
   \underbrace{ \mathbf{U}_s^{-1}\mathbf{T}^n \mathbf{U}_s}_{\mathbf{H}_s} 
   \begin{Bmatrix}
    a_1 \\ a_2
    \end{Bmatrix} = \lambda^n
    \begin{Bmatrix}
    a_1 \\ a_2
    \end{Bmatrix}.
    \label{eq:EVP_H}
\end{equation}
where the matrix $\mathbf{H}_s$ is given by
\begin{equation}
    \mathbf{H}_s = 
    \begin{bmatrix}
    h & \hat{h} \\ \hat{h}^\dagger & h^{\dagger}
    \end{bmatrix}.
\end{equation}
where the superscript $\dagger$ denotes the complex conjugate and
\begin{subequations}
\begin{equation}
    h = \frac{1}{2} (t_{11n} + t_{22n}) + \frac{\textbf{i}}{2}\left( \frac{t_{21n}}{E_s A_s k_s} - (E_s A_s k_s) t_{12n} \right),
\end{equation}
\begin{equation}
    \hat{h} = \frac{1}{2} ( t_{11n} - t_{22n}) + \frac{\textbf{i}}{2}\left( \frac{t_{21n}}{E_s A_s k_s} + (E_s A_s k_s) t_{12n} \right).
\end{equation}
\end{subequations}
Considering the first row of Eq.~(\ref{eq:EVP_H}) in conjunction with $a_2 = a_1$ from the zero force condition at $x = 0$ pertaining to a free end, we obtain
\begin{equation}
    t_{11n} - \lambda^n + \textbf{i}\left( \frac{t_{21n}}{E_s A_s k_s} \right) = 0.
    \label{eq:EVP_1}
\end{equation}
Equivalently, the second row of Eq.~(\ref{eq:EVP_H}) can be used and shall result in the complex conjugate of Eq.~(\ref{eq:EVP_H}). Note that both real and imaginary components have to be zero to satisfy Eq.~(\ref{eq:EVP_1}). Recall that, for a natural frequency within the bandgap, the eigenvalues of the transfer matrix are both real and distinct, and only $|\lambda|< 1$ shall be considered for a mode decaying away from $x = 0$, hence, $\lambda^n \rightarrow 0$ as $n \rightarrow \infty$. As a consequence, the result of setting the real and imaginary components of Eq.~(\ref{eq:EVP_1}) equal to zero becomes merely the characteristic equations of the free-fixed ($t_{11n} = 0$) and free-free ($t_{21n} = 0$) cases, respectively. This signals that a truncation resonance must satisfy both the characteristic equations of the free-fixed and free-free cases, which confirms that these modes are in fact the same for both boundary conditions when $n\rightarrow \infty$. If instead we consider a fixed-end scenario (i.e, $u_0 = 0$ and $a_2 = -a_1$), it can be similarly shown that
\begin{equation}
    t_{22n} - \lambda^n + \textbf{i}\left(E_s A_s k_s \right) t_{12n} = 0,
    \label{eq:EVP_2}
\end{equation}
and hence $t_{22n} = 0$ and $t_{12n} = 0$ must both be satisfied when $\lambda^n \rightarrow 0$, which serves as a proof for the truncation resonance matching between the free-fixed and fixed-fixed cases.

\subsection{Effect of unit cell symmetry on truncation resonances}
\label{sec:effect_of_symmetry}

\begin{figure*}[]
     \centering
\includegraphics[width=\textwidth]{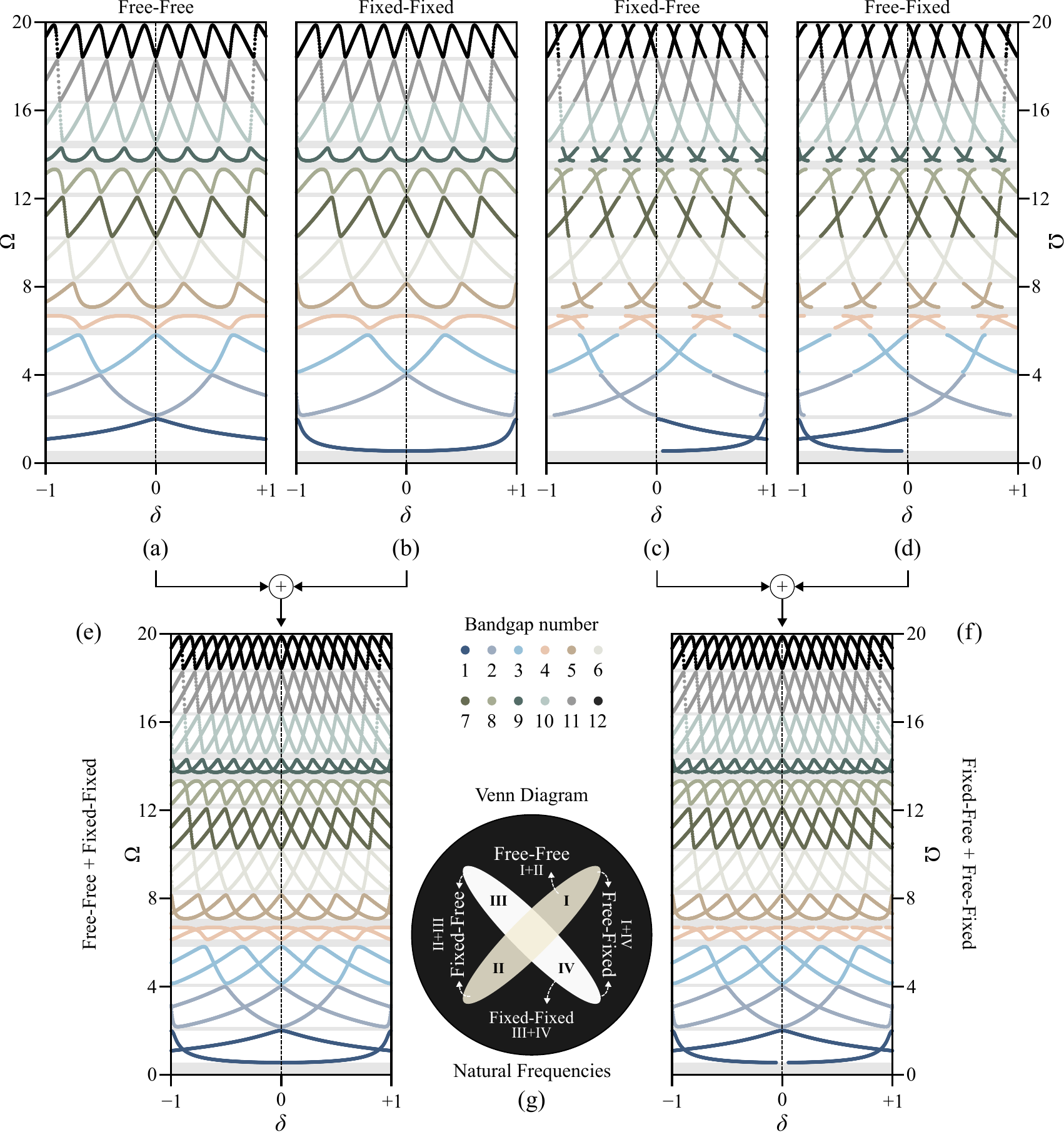}
     \caption{Effect of the symmetry parameter $\delta$ on the truncation resonances of finite PnC rods subject to the following boundary conditions: (a) Free-free, (b) Fixed-fixed, (c) Fixed-free, and (d) Free-fixed. The combined truncation resonances of the free-free and fixed-fixed, and the combined truncation resonances of the fixed-free and free-fixed are shown in (e) and (f), respectively. (g) Venn diagram summarizing the distribution of truncation resonances across the four boundary conditions. The outer black circle represents all possible natural frequencies of the finite PnC rod subject to any boundary condition. The two ellipses inside the black circle represent the entire set of truncation resonances. The space outside of the two ellipses contains the remaining natural frequencies representing pass band resonances. Groups I and II represents the free-free truncation resonances and groups III and IV represents the fixed-fixed truncation resonances. The shared area (intersection) between the two ellipses contains truncation resonances which co-exist in both cases. The combination (union) of groups II and III constitute the truncation resonances for the fixed-free case, while the combination (union) of I and IV constitute the truncation resonances for the free-fixed case.}
     \label{fig:Effect_of_Symmetry_Delta}
\end{figure*}

The performance of the finite PnC rods shown thus far in Figs.~\ref{fig:FRF} through \ref{fig:mode_shapes} is for $\delta = -0.8$. However, the choice of unit cell (i.e., symmetry parameter) can largely influence the final configuration of the finite structure and thus the location of the truncation resonances, as will be outlined here. Figure~\ref{fig:Effect_of_Symmetry_Delta} illustrates the effect of $\delta$ on the truncation resonance frequencies. The plots show a series of loci which trace each truncation frequency (color coded to represent the bandgap to which they belong) as $\delta$ varies from $-1$ to $1$. This is done for all four boundary conditions discussed in Sec.~\ref{sec:Nat_Freq_BCs}. Since a truncation resonance can be located anywhere between the starting and ending frequencies of the bandgap in which it resides, the shaded space between the different curves in Fig.~\ref{fig:Effect_of_Symmetry_Delta} actually represents pass band regions. As can be seen in the figure, the variation of truncation resonance location as the unit cell choice changes fluctuates up and down in an oscillatory manner. More fluctuations occur at larger values of $\Omega$ indicating an increased sensitivity to the symmetry parameter $\delta$ at higher frequencies.

Next, we shift our focus to the combined effect of $\delta$ and boundary conditions on truncation resonance locations. The first observation that stands out is the symmetry of the curves in Figs.~\ref{fig:Effect_of_Symmetry_Delta}(a,b), representing the free-free and fixed-fixed rods, about $\delta=0$. This is understandable given that the boundary condition is the same on both ends of the structure in both of these cases, and thus a change in the sign of $\delta$ would not result in a change in the system's truncation resonances. This has also been mathematically confirmed in Eqs.~(\ref{eq:possible_TP}) and (\ref{eq:possible_TP2}) where $\delta$ appears as a factor inside the cosine function, where the sign of $\delta$ would not affect the function's outcome. On the other hand, the truncation resonance curves are not symmetric about $\delta=0$ for the fixed-free and free-fixed cases, as shown in Figs.~\ref{fig:Effect_of_Symmetry_Delta}(c,d). This is anticipated due to the asymmetric nature of the rod under these two boundary conditions. As established in Sec.~\ref{sec:Fixed_Free_trunc}, each truncation resonance in these two cases must have an identical counterpart in either the free-free or fixed-fixed rods. For that reason, it can be observed that parts of the curves shown in Figs.~\ref{fig:Effect_of_Symmetry_Delta}(a,b) appear in Figs.~\ref{fig:Effect_of_Symmetry_Delta}(c,d). An interesting graphical depiction of this phenomenon is shown in Figs.~\ref{fig:Effect_of_Symmetry_Delta}(e,f) which show that the combined truncation resonances of the free-free and fixed-fixed cases are nearly identical to the combined truncation resonances of the fixed-free and free-fixed ones. To further elucidate the overall patterns shown in Fig.~\ref{fig:Effect_of_Symmetry_Delta}, we break down the truncation resonances of the free-free case into two groups. The first represents mode shapes with localized vibrations near $x=0$ and is referred to as group I, while the second represents mode shapes with localized vibrations near $x = n\ell$ and is referred to as group II. Similarly, we break down the truncation resonances of the fixed-fixed case into groups III and IV. With these four groups in mind, the Venn diagram shown in Fig.~\ref{fig:Effect_of_Symmetry_Delta}(g) summarizes the distribution of truncation resonances across the four boundary conditions. The outer black circle represents all the possible natural frequencies of the finite PnC rod within a given frequency range, subject to any of the four boundary conditions. The two ellipses inside the black circle represent the entire set of truncation resonances in that frequency range, while the space outside of the two ellipses contains the remaining natural frequencies representing pass band resonances. The ellipse comprising groups I and II represents the free-free truncation resonances and the ellipse comprising groups III and IV represents the fixed-fixed truncation resonances. The shared area (intersection) between the two ellipses contains some truncation resonances which co-exist in both cases, as evident from the crossing of curves in Figs.~\ref{fig:Effect_of_Symmetry_Delta}(e,f). Finally, the combination (union) of groups II and III constitute the truncation resonances for the fixed-free case, while the combination (union) of I and IV constitute the truncation resonances for the free-fixed case.

\newpage

\begin{figure*}[h]
     \centering
\includegraphics[width=\textwidth]{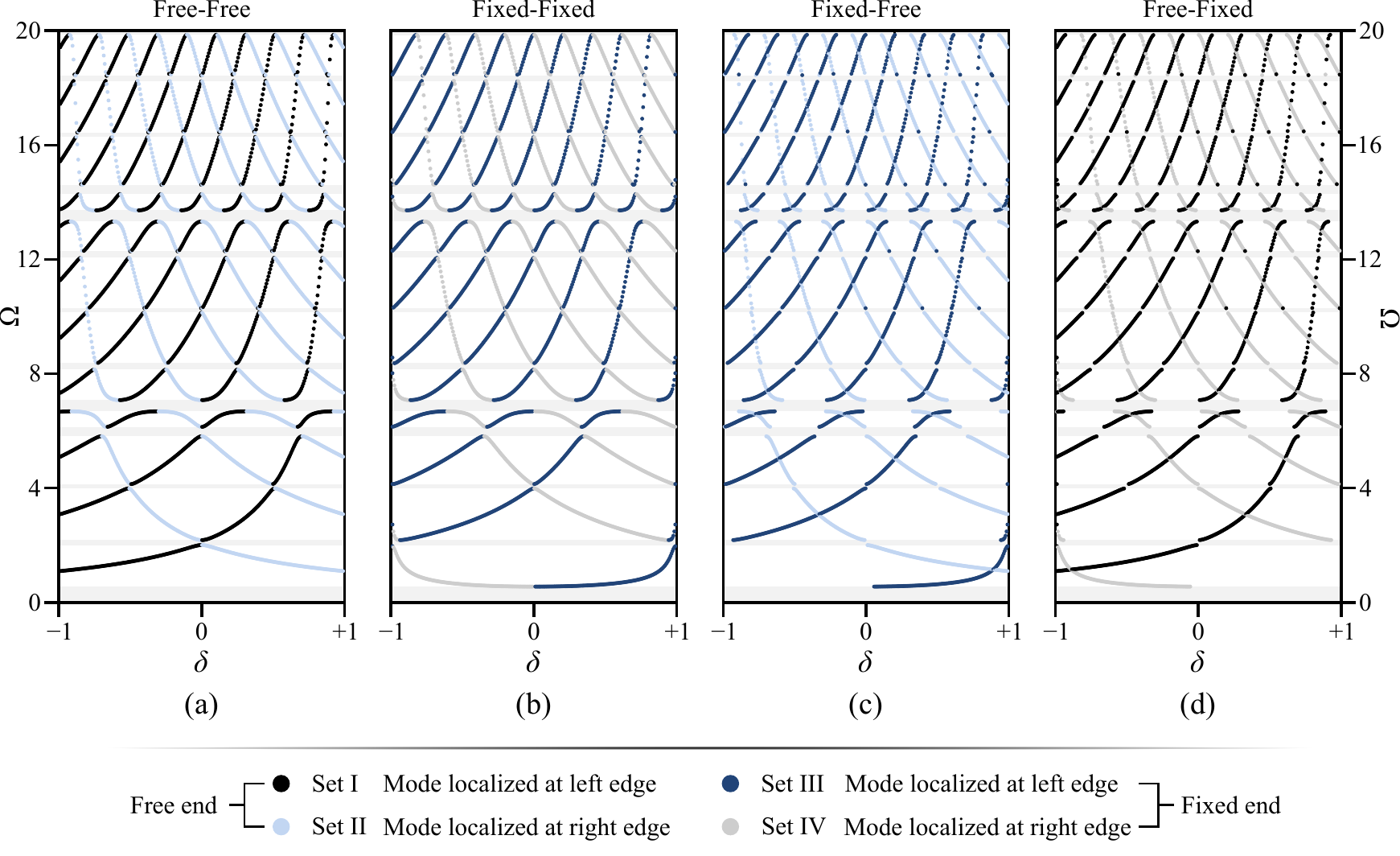}
     \caption{Mode shapes corresponding to the truncation resonances of the finite PnC rod are classified into four sets as indicated above depending on the edge where localized vibrations take place and the type of boundary condition. Following this classification, truncation resonances from Figs.~\ref{fig:Effect_of_Symmetry_Delta}(a-d) are re-plotted here versus the symmetry parameter $\delta$ for the following boundary conditions: (a) Free-free, (b) Fixed-fixed, (c) Fixed-free, and (d) Free-fixed.}
     \label{fig:localization}
\end{figure*}

To further emphasize the truncation resonance features summarized in Fig.~\ref{fig:Effect_of_Symmetry_Delta}(g), the eigenvectors (i.e., mode shapes) corresponding to the truncation resonances are computed using our FEM model previously described in Sec.~\ref{sec:FRF}. These modes are then categorized based on the edge of the structure at which the vibrations are localized. This is accomplished by calculating a normalized weighted average $\bar{x}$ of the location using the absolute displacements of the FEM nodes as weights, denoted as $u_j$, and their location in the spatial domain as the data, denoted as $x_j$, which is mathematically expressed as follows
\begin{equation}
        \bar{x} = \frac{1}{n\ell}\frac{\sum_j x_j |u_j|}{\sum_j |u_j|}.
        \label{eq:avg_location}
\end{equation}
A normalized average of $\bar{x}<0.50$ or $\bar{x}>0.50$ indicates that higher displacement amplitudes are more skewed towards the left or right edges, respectively. The limiting case of $\bar{x}\rightarrow0$ or $\bar{x}\rightarrow1$ indicates that the amplitude is strongly localized at the left or right edge, respectively. Figure~\ref{fig:localization} shows the outcome of this procedure for all four boundary conditions, namely, (a) free-free, (b) fixed-fixed, (c) fixed-free, and (d) free-fixed for all unit cell choices spanning the symmetry parameter range $-1 \leq \delta \leq 1$. For ease of reference, the modes in the figure are color coded to reflect the edge at which the localized vibrations take place (left vs. right) as well as the boundary condition at that edge (free vs. fixed), and are therefore grouped in four sets as detailed in the figure's legend. As seen in Figs.~\ref{fig:localization}(a,b), the truncation resonances in both the free-free and fixed-fixed cases consist of left- and right-localized modes, represented by sets I + II and III + IV, respectively. Additionally, it can be observed that sets II + III constitute the truncation resonances of the fixed-free case in Fig.~\ref{fig:localization}(c). This makes perfect sense since these sets represent the matching boundary conditions at the two respective edges. Specifically, the eigenvectors in set II represent ``right-localized modes for a free end" while the eigenvectors in set III represent ``left-localized modes for a fixed end". These two sets, put together, create the fixed-free PnC. Following the same logic, it is observed that sets I + IV constitute the truncation resonances of the free-fixed case in Fig.~\ref{fig:localization}(d). Lastly, we make an important observation from Figs.~\ref{fig:localization}(a,b) regarding a truncation resonance phenomenon which is not immediately evident from Figs.~\ref{fig:Effect_of_Symmetry_Delta}(a,b), which is that the edge at which the localized vibrations happen flips from left to right or vice versa upon touching a pass band (remember that all these truncation resonances are by definition inside bandgaps). It should be noted that a similar behavior has been previously reported in the context of quasi-periodic lattices~\cite{Rosa2019EdgeLattices}.

\subsection{Absence of truncation resonances in symmetric unit cells ($\delta = 0$)}

A perfectly symmetric unit cell is a special case for all boundary conditions, where truncation resonances cease to exist. To analytically prove this, we start with the free-fixed and fixed-free cases whose characteristic equations are given by $t_{11n} = 0$ and $t_{22n} = 0$, respectively. For $\delta=0$, both of these reduce to
\begin{equation}
    \cos(nq) = 0,
\end{equation}
for which the only possible wavenumber solutions are real and are given by
\begin{equation}
    q = \frac{(2r-1)\pi}{2n}, \ r = 1,2,\dots n.
\end{equation}
All of these real wavenumbers correspond to natural frequencies that fall within pass bands. These natural frequencies can simply be found by solving the dispersion relation in Eq.~(\ref{eq:non-dim_disp_rel}) at these wavenumbers, thereby proving that no truncation resonances shall appear for these two boundary conditions as long as $\delta=0$. 

While no truncation resonances exist for the free-free and fixed-fixed boundary conditions when $\delta=0$, proving so is not as straightforward. In these two cases, the portion of the characteristic equations responsible for truncation resonances, i.e., Eqs.~(\ref{eq:possible_TP}) and (\ref{eq:possible_TP2}) reduce to
\begin{equation}
    \Big(\cos(\Omega) \mp \beta \cos(\Omega \alpha)\Big)\Big(\sin(\Omega) \pm \beta\sin(\Omega \alpha) \Big) = 0,
    \label{eq:Ch_eq_freefree_delta0}
\end{equation}
for free-free $(-,+)$ and fixed-fixed $(+,-)$ boundary conditions. Both of the bracketed expressions in Eq.~(\ref{eq:Ch_eq_freefree_delta0}) are a multiplication of two terms, which are in fact related to the equations that govern the bandgap limits. To elaborate, let's substitute $q = 0$ and $q=\pi$, the two values which mark the beginning and end of all dispersion branches, in the dispersion relation given by Eq.~(\ref{eq:non-dim_disp_rel}), which results in the following
\begin{equation}
    \cos(2\Omega) - \beta^2 \cos(2\Omega \alpha) \pm (1-\beta^2) = 0.
    \label{eq:odd_BG_Eq_1}
\end{equation}
Equation~(\ref{eq:odd_BG_Eq_1}) provides the limits for odd $(+)$ and even $(-)$ numbered bandgaps, and can be further simplified to
\begin{subequations}
\begin{equation}
   \Big( \cos(\Omega) - \beta \cos(\Omega \alpha)\Big)\Big( \cos(\Omega) + \beta \cos(\Omega \alpha)\Big) = 0.
    \label{eq:BG_limit_2}
\end{equation}
\begin{equation}
    \Big(\sin(\Omega) - \beta\sin(\Omega \alpha) \Big)\Big(\sin(\Omega) + \beta\sin(\Omega \alpha) \Big) = 0,
    \label{eq:BG_limit_1}
\end{equation}
\end{subequations}
One can now see the connection between the equations defining bandgap limits and the characteristic equations of the free-free and fixed-fixed cases when $\delta = 0$. Specifically, the two equations in Eq.~(\ref{eq:Ch_eq_freefree_delta0}) can be simply obtained by multiplying one term in Eq.~(\ref{eq:BG_limit_2}) by another in Eq.~(\ref{eq:BG_limit_1}). This effectively proves that all resonances corresponding to these two boundary conditions when $\delta=0$ are indeed within pass bands. Similar results have been reported by Mead for mono-coupled systems, where some natural frequencies for asymmetric elements arise within attenuation zones and coincide with bounding frequencies of bandgaps for symmetric elements \cite{mead1975wave}. The same phenomenon has also been observed in some configurations of diatomic lattices with free boundaries \cite{al2017pole}.

\begin{figure*}[h]
     \centering
\includegraphics[]{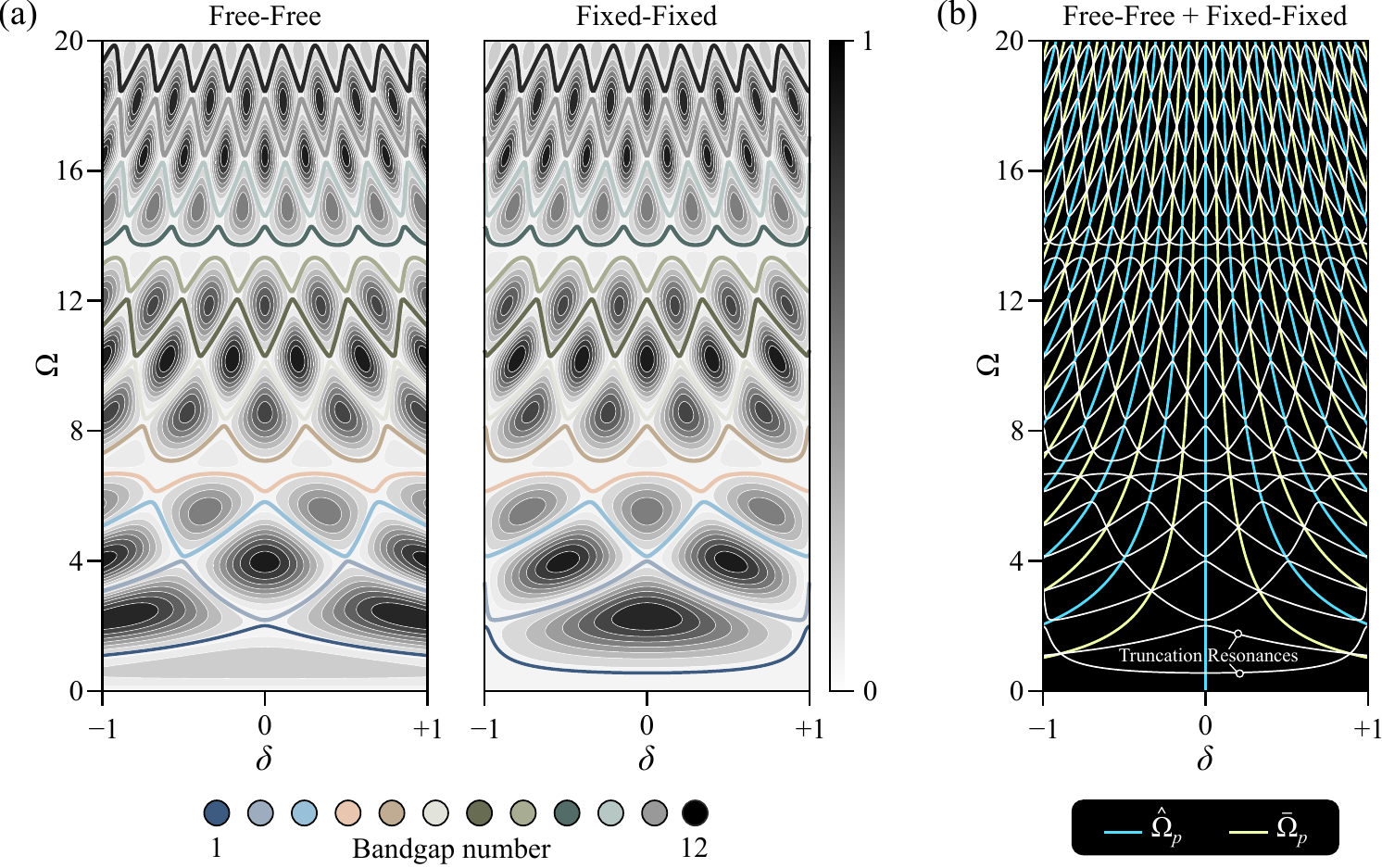}
     \caption{(a) Relationship between the truncation resonances patterns and the contours of Eqs.~(\ref{eq:possible_TP}) and (\ref{eq:possible_TP2}). Color-coded solid lines between the contours represent points that render the contour's magnitude equal to zero, i.e., roots of Eqs.~(\ref{eq:possible_TP}) and (\ref{eq:possible_TP2}). (b) The combined patterns of the free-free and fixed-fixed truncation resonances, given by the lines described in Eqs.~(\ref{eq:cosine1}) and (\ref{eq:cosine0}).}
     \label{fig:Pattern_Of_Solution}
\end{figure*}

\begin{figure*}[]
     \centering
\includegraphics[]{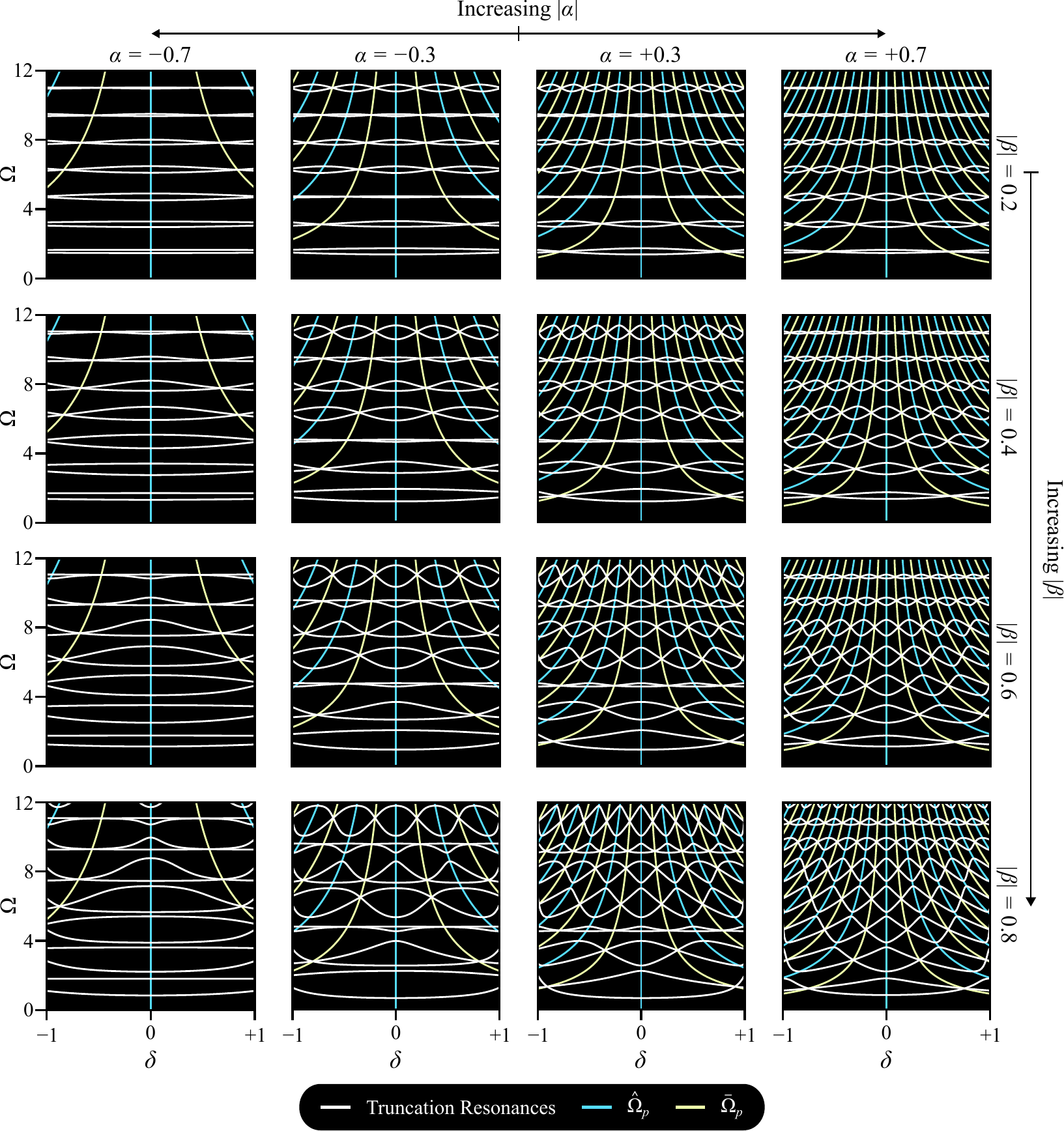}
     \caption{Effect of system parameters, namely the frequency contrast $\alpha$ and the impedance contrast $\beta$, on the overall patterns of the truncation resonances across all the possible unit cell configurations, as dictated by the full $\delta$ range.}
     \label{fig:Effect_of_Alpha_Beta}
\end{figure*}

\subsection{Patterns of truncation resonances}
\label{sec:pattern_study}

The general pattern of the truncation resonance curves is best explained using Eqs.~(\ref{eq:possible_TP}) and (\ref{eq:possible_TP2}). For given values of $\beta$ and $\alpha$, a contour plot can be obtained by sweeping $\Omega$ and $\delta$, as can be seen in Fig.~\ref{fig:Pattern_Of_Solution}(a) for the free-free and fixed-fixed cases. Whenever a contour has a magnitude of zero, it represents a solution of the equation (i.e., a truncation resonance), which are indicated by the color-coded solid curves in the figure. The proof presented earlier confirming the absence of truncation resonance for free-free and fixed-fixed PnCs which are formed of symmetric unit cells (i.e., when $\delta=0$) can be used to explain the general patterns of the truncation resonances as a function of $\delta$. We start by pointing out that the magnitude of the cosine function $\cos\big(\delta(1+\alpha)\Omega\big)$ in Eqs.~(\ref{eq:possible_TP}) and (\ref{eq:possible_TP2}) becomes unity when $\delta = 0$ and results in the simplified characteristic equations in Eq.~(\ref{eq:Ch_eq_freefree_delta0}). Second, given its periodicity, the cosine function may assume the values of $\pm 1$ at different combinations of $\Omega$ and $\delta$. As such, one can derive an expression that relates the frequency $\Omega$ to the symmetry parameter $\delta$ that satisfies such condition, which results in
\begin{equation}
    \hat{\Omega}_p = \frac{\pm p\pi}{(1+\alpha)} \frac{1}{\delta},
    \label{eq:cosine1}
\end{equation}
where $p = 0, 1, 2, \dots, \infty$. These lines intersect with the truncation resonance curves when they touch the bandgap limits, as observed in Fig.~\ref{fig:Pattern_Of_Solution}(b). In addition, it is possible that the magnitude of $\cos(\delta (1+\alpha)\Omega)$ becomes zero at some combinations of $\Omega$ and $\delta$, resulting in
\begin{equation}
    \bar{\Omega}_p = \frac{\pm (2p+1)\pi}{2(1+\alpha)} \frac{1}{\delta}.
    \label{eq:cosine0}
\end{equation}
In such a case, the curves for free-free and fixed-fixed truncation resonances intersect since $t_{12n} = 0$ and $t_{21n} = 0$ have identical roots, and the intersection point satisfies Eq.~(\ref{eq:cosine0}). This also aligns with the shared area between the two ellipses representing the free-free and fixed-fixed truncation resonances in the Venn diagram of Fig.~\ref{fig:Effect_of_Symmetry_Delta}(g).

Finally, for completeness, we study the effect of the design parameters on the truncation resonance curves. As depicted in Fig.~\ref{fig:Pattern_Of_Solution}, such curves fluctuate in an oscillatory manner as the value of $\delta$ changes, and become more sensitive to changes in $\delta$ at higher frequencies. Figure~\ref{fig:Effect_of_Alpha_Beta} shows the combined truncation resonances for the free-free and fixed-fixed cases for different values of $\alpha$ (columns) and $\beta$ (rows), from which the following can be established: 

\begin{itemize}
    
    \item Equations~(\ref{eq:cosine1}) predicts when the solutions of the free-free and fixed-fixed truncation resonances touch the bandgap limits, while Eq.~(\ref{eq:cosine0}) predict when the solutions of the free-free and fixed-fixed truncation resonances intersect with one another, as established earlier.
    
    \item The frequency of the oscillatory behavior of the truncation resonance curves depends on $\alpha$. It is seen that a move towards increasingly negative values of $\alpha$ decreases the frequency by which truncation resonances fluctuate in response to a change in $\delta$ over the same $\Omega$ range. This behavior can be well explained by inspecting the lines generated by Eq.~(\ref{eq:cosine0}), which captures all the intersections between truncation resonance lines, as shown in Fig.~\ref{fig:Effect_of_Alpha_Beta}. The first solution of this equation becomes larger with a smaller denominator due to the $(1+\alpha)$ term. As a result, integer multiples of the first solution grow further apart, thus forcing truncation resonance curves to oscillate a fewer number of times over the $\delta$ range.
    
    \item The impedance contrast $\beta$ plays no role in the periodic fluctuations of the truncation resonance curves as $\delta$ changes. This can be inferred from Eqs.~(\ref{eq:cosine1})~and~(\ref{eq:cosine0}) since they are not functions of $\beta$. However, the amplitude of the fluctuation is largely a function $\beta$ and is bounded by the width of the underlying bandgap. 
    
    \item The sign of $\beta$ does not affect the behavior of the truncation resonances curves when we collectively consider the free-free and fixed-fixed cases, owing to the fact that flipping the sign of $\beta$ only interchanges the curves of the free-free and fixed-fixed cases, as evident in Eqs.~(\ref{eq:possible_TP})~and~(\ref{eq:possible_TP2}).
\end{itemize}

\section{Conclusions}

In this paper, we presented a comprehensive analytical investigation of truncation resonances emerging within the bandgaps of continuum rod-based finite PnCs. By implementing an approach based on the transfer matrix method (TMM), expressions for both the unit cell dispersion relation of the infinite PnC rod and the frequency response functions of a finite one with defined boundaries were obtained. Different truncated chains were obtained by stacking a number of unit cells, each consisting of two different material/geometric segments and a level of symmetry characterized by the parameter $\delta$. It was proven that the truncation resonances of a free-free and fixed-fixed PnC rod always coincide with the natural frequencies of a single unit cell with the same boundary conditions, providing an efficient tool to tune these truncation resonances. Furthermore, it was shown that the truncation resonances exhibited by the two other cases (fixed-free and free-fixed) must co-exist in the two former cases (free-free and fixed-fixed). This feature was further verified and analytically proven using a semi-infinite PnC configuration. With the exception of the perfectly symmetric unit cell choice (i.e., $\delta=0$), it was also shown that truncation resonances can potentially occur in all the commonly encountered boundary conditions and are functions of the symmetry parameter $\delta$ as well as the unit cell properties, defined here using the frequency and impedance contrasts, $\alpha$ and $\beta$. In the free-free and fixed-fixed cases, a maximum of one truncation resonance may occur inside each bandgap, while in the fixed-free and free-fixed cases, up to two truncation resonances may emerge inside each bandgap. The underlying connection between the localized mode associated with a truncation resonance and the corresponding boundary condition was explained, enabling us to interpret the different patterns of truncation resonances for a family of finite PnCs obtained using different asymmetric realizations of a single unit cell. The ability to understand the origins of truncation resonances beyond lumped-parameter systems, and to accurately predict these resonances in continuum rod-based PnCs provides an intriguing road map for a broad range of applications that demonstrably benefit from the presence of truncation resonances.

\section*{Funding}

M. Nouh acknowledges support of this work by the US National Science Foundation through research award no. 1847254 (CAREER) and 1904254. A. T. Juhl acknowledges support of this work by the Air Force Office of Scientific Research via award no. 20RXCOR058.

\section*{Data availability}

All the data used and/or analysed during the current study is
included in the manuscript. 

\bibliographystyle{model1-num-names}
\bibliography{references.bib}

\begin{thebibliography}{48}
\expandafter\ifx\csname natexlab\endcsname\relax\def\natexlab#1{#1}\fi
\providecommand{\bibinfo}[2]{#2}
\ifx\xfnm\relax \def\xfnm[#1]{\unskip,\space#1}\fi
\bibitem[{Vasileiadis et~al.(2021)Vasileiadis, Varghese, Babacic, Gomis-Bresco,
  Navarro~Urrios, and Graczykowski}]{vasileiadis2021progress}
\bibinfo{author}{T.~Vasileiadis}, \bibinfo{author}{J.~Varghese},
  \bibinfo{author}{V.~Babacic}, \bibinfo{author}{J.~Gomis-Bresco},
  \bibinfo{author}{D.~Navarro~Urrios}, \bibinfo{author}{B.~Graczykowski},
\newblock \bibinfo{title}{Progress and perspectives on phononic crystals},
\newblock \bibinfo{journal}{Journal of Applied Physics} \bibinfo{volume}{129}
  (\bibinfo{year}{2021}) \bibinfo{pages}{160901}.
\bibitem[{Dalela et~al.(2022)Dalela, Balaji, and Jena}]{dalela2022review}
\bibinfo{author}{S.~Dalela}, \bibinfo{author}{P.~Balaji},
  \bibinfo{author}{D.~Jena},
\newblock \bibinfo{title}{A review on application of mechanical metamaterials
  for vibration control},
\newblock \bibinfo{journal}{Mechanics of advanced materials and structures}
  \bibinfo{volume}{29} (\bibinfo{year}{2022}) \bibinfo{pages}{3237--3262}.
\bibitem[{Al~Ba'Ba'A and Nouh(2017)}]{AlBabaa2016a}
\bibinfo{author}{H.~Al~Ba'Ba'A}, \bibinfo{author}{M.~Nouh},
\newblock \bibinfo{title}{{An investigation of vibrational power flow in
  one-dimensional dissipative phononic structures}},
\newblock \bibinfo{journal}{Journal of Vibration and Acoustics, Transactions of
  the ASME} \bibinfo{volume}{139} (\bibinfo{year}{2017})
  \bibinfo{pages}{21003--21010}.
\bibitem[{Liu and Hussein(2012)}]{liu2012wave}
\bibinfo{author}{L.~Liu}, \bibinfo{author}{M.~I. Hussein},
\newblock \bibinfo{title}{Wave motion in periodic flexural beams and
  characterization of the transition between bragg scattering and local
  resonance},
\newblock \bibinfo{journal}{Journal of Applied Mechanics} \bibinfo{volume}{79}
  (\bibinfo{year}{2012}).
\bibitem[{Foehr et~al.(2018)Foehr, Bilal, Huber, and
  Daraio}]{Foehr2018Spiral-BasedInsulators}
\bibinfo{author}{A.~Foehr}, \bibinfo{author}{O.~R. Bilal},
  \bibinfo{author}{S.~D. Huber}, \bibinfo{author}{C.~Daraio},
\newblock \bibinfo{title}{{Spiral-Based Phononic Plates: From Wave Beaming to
  Topological Insulators}},
\newblock \bibinfo{journal}{Physical Review Letters} \bibinfo{volume}{120}
  (\bibinfo{year}{2018}) \bibinfo{pages}{205501}.
\bibitem[{Hussein et~al.(2014)Hussein, Leamy, and Ruzzene}]{Hussein2014}
\bibinfo{author}{M.~I. Hussein}, \bibinfo{author}{M.~J. Leamy},
  \bibinfo{author}{M.~Ruzzene},
\newblock \bibinfo{title}{{Dynamics of phononic materials and structures:
  Historical origins, recent progress, and future outlook}},
\newblock \bibinfo{journal}{Applied Mechanics Reviews} \bibinfo{volume}{66}
  (\bibinfo{year}{2014}) \bibinfo{pages}{040802}.
\bibitem[{Tanaka et~al.(2000)Tanaka, Tomoyasu, and Tamura}]{tanaka2000band}
\bibinfo{author}{Y.~Tanaka}, \bibinfo{author}{Y.~Tomoyasu},
  \bibinfo{author}{S.-i. Tamura},
\newblock \bibinfo{title}{Band structure of acoustic waves in phononic
  lattices: Two-dimensional composites with large acoustic mismatch},
\newblock \bibinfo{journal}{Physical Review B} \bibinfo{volume}{62}
  (\bibinfo{year}{2000}) \bibinfo{pages}{7387}.
\bibitem[{Hu et~al.(2021)Hu, Tang, Liang, Lan, and Das}]{hu2021acoustic}
\bibinfo{author}{G.~Hu}, \bibinfo{author}{L.~Tang}, \bibinfo{author}{J.~Liang},
  \bibinfo{author}{C.~Lan}, \bibinfo{author}{R.~Das},
\newblock \bibinfo{title}{Acoustic-elastic metamaterials and phononic crystals
  for energy harvesting: A review},
\newblock \bibinfo{journal}{Smart Materials and Structures}
  (\bibinfo{year}{2021}).
\bibitem[{Lin et~al.(2021)Lin, Al~Ba’ba’a, and Tol}]{lin2021piezoelectric}
\bibinfo{author}{Z.~Lin}, \bibinfo{author}{H.~Al~Ba’ba’a},
  \bibinfo{author}{S.~Tol},
\newblock \bibinfo{title}{Piezoelectric metastructures for simultaneous
  broadband energy harvesting and vibration suppression of traveling waves},
\newblock \bibinfo{journal}{Smart Materials and Structures}
  \bibinfo{volume}{30} (\bibinfo{year}{2021}) \bibinfo{pages}{075037}.
\bibitem[{Bilal et~al.(2018)Bilal, Ballagi, and Daraio}]{bilal2018architected}
\bibinfo{author}{O.~R. Bilal}, \bibinfo{author}{D.~Ballagi},
  \bibinfo{author}{C.~Daraio},
\newblock \bibinfo{title}{{Architected Lattices for Simultaneous Broadband
  Attenuation of Airborne Sound and Mechanical Vibrations in All Directions}},
\newblock \bibinfo{journal}{Physical Review Applied} \bibinfo{volume}{10}
  (\bibinfo{year}{2018}) \bibinfo{pages}{54060}.
\bibitem[{Attarzadeh and Nouh(2018{\natexlab{a}})}]{attarzadeh2018non}
\bibinfo{author}{M.~A. Attarzadeh}, \bibinfo{author}{M.~Nouh},
\newblock \bibinfo{title}{{Non-reciprocal elastic wave propagation in 2D
  phononic membranes with spatiotemporally varying material properties}},
\newblock \bibinfo{journal}{Journal of Sound and Vibration}
  \bibinfo{volume}{422} (\bibinfo{year}{2018}{\natexlab{a}})
  \bibinfo{pages}{264--277}.
\bibitem[{Attarzadeh and Nouh(2018{\natexlab{b}})}]{attarzadeh2018elastic}
\bibinfo{author}{M.~A. Attarzadeh}, \bibinfo{author}{M.~Nouh},
\newblock \bibinfo{title}{{Elastic wave propagation in moving phononic crystals
  and correlations with stationary spatiotemporally modulated systems}},
\newblock \bibinfo{journal}{AIP Advances} \bibinfo{volume}{8}
  (\bibinfo{year}{2018}{\natexlab{b}}) \bibinfo{pages}{105302}.
\bibitem[{Danawe et~al.(2020)Danawe, Okudan, Ozevin, and
  Tol}]{danawe2020conformal}
\bibinfo{author}{H.~Danawe}, \bibinfo{author}{G.~Okudan},
  \bibinfo{author}{D.~Ozevin}, \bibinfo{author}{S.~Tol},
\newblock \bibinfo{title}{Conformal gradient-index phononic crystal lens for
  ultrasonic wave focusing in pipe-like structures},
\newblock \bibinfo{journal}{Applied Physics Letters} \bibinfo{volume}{117}
  (\bibinfo{year}{2020}) \bibinfo{pages}{021906}.
\bibitem[{Bloch(1929)}]{Bloch1929}
\bibinfo{author}{F.~Bloch},
\newblock \bibinfo{title}{{{\"{U}}ber die Quantenmechanik der Elektronen in
  Kristallgittern}},
\newblock \bibinfo{journal}{Zeitschrift f{\"{u}}r Physik} \bibinfo{volume}{52}
  (\bibinfo{year}{1929}) \bibinfo{pages}{555--600}.
\bibitem[{Al~Ba'ba'a et~al.(2019)Al~Ba'ba'a, Nouh, and
  Singh}]{AlBabaa2019DispersionCrystals}
\bibinfo{author}{H.~Al~Ba'ba'a}, \bibinfo{author}{M.~Nouh},
  \bibinfo{author}{T.~Singh},
\newblock \bibinfo{title}{{Dispersion and topological characteristics of
  permutative polyatomic phononic crystals}},
\newblock \bibinfo{journal}{Proceedings of the Royal Society A: Mathematical,
  Physical and Engineering Sciences} \bibinfo{volume}{475}
  (\bibinfo{year}{2019}).
\bibitem[{Davis et~al.(2011)Davis, Tomchek, Flores, Liu, and
  Hussein}]{davis2011analysis}
\bibinfo{author}{B.~L. Davis}, \bibinfo{author}{A.~S. Tomchek},
  \bibinfo{author}{E.~A. Flores}, \bibinfo{author}{L.~Liu},
  \bibinfo{author}{M.~I. Hussein},
\newblock \bibinfo{title}{Analysis of periodicity termination in phononic
  crystals},
\newblock in: \bibinfo{booktitle}{ASME International Mechanical Engineering
  Congress and Exposition}, volume \bibinfo{volume}{54945}, pp.
  \bibinfo{pages}{973--977}.
\bibitem[{Roman and Sebastian(2015)}]{susstrunk2015observation}
\bibinfo{author}{S.~Roman}, \bibinfo{author}{D.~H. Sebastian},
\newblock \bibinfo{title}{{Observation of phononic helical edge states in a
  mechanical topological insulator}},
\newblock \bibinfo{journal}{Science} \bibinfo{volume}{349}
  (\bibinfo{year}{2015}) \bibinfo{pages}{47--50}.
\bibitem[{Prodan et~al.(2017)Prodan, Dobiszewski, Kanwal, Palmieri, and
  Prodan}]{Prodan2017DynamicalSystems}
\bibinfo{author}{E.~Prodan}, \bibinfo{author}{K.~Dobiszewski},
  \bibinfo{author}{A.~Kanwal}, \bibinfo{author}{J.~Palmieri},
  \bibinfo{author}{C.~Prodan},
\newblock \bibinfo{title}{{Dynamical Majorana edge modes in a broad class of
  topological mechanical systems}},
\newblock \bibinfo{journal}{Nature Communications} \bibinfo{volume}{8}
  (\bibinfo{year}{2017}).
\bibitem[{Rosa et~al.(2019)Rosa, Pal, Arruda, and
  Ruzzene}]{Rosa2019EdgeLattices}
\bibinfo{author}{M.~I. Rosa}, \bibinfo{author}{R.~K. Pal},
  \bibinfo{author}{J.~R. Arruda}, \bibinfo{author}{M.~Ruzzene},
\newblock \bibinfo{title}{{Edge States and Topological Pumping in Spatially
  Modulated Elastic Lattices}},
\newblock \bibinfo{journal}{Physical Review Letters} \bibinfo{volume}{123}
  (\bibinfo{year}{2019}) \bibinfo{pages}{034301}.
\bibitem[{Hussein et~al.(2015)Hussein, Biringen, Bilal, and
  Kucala}]{hussein2015flow}
\bibinfo{author}{M.~I. Hussein}, \bibinfo{author}{S.~Biringen},
  \bibinfo{author}{O.~R. Bilal}, \bibinfo{author}{A.~Kucala},
\newblock \bibinfo{title}{{Flow stabilization by subsurface phonons}},
\newblock \bibinfo{journal}{Proceedings of the Royal Society A: Mathematical,
  Physical and Engineering Sciences} \bibinfo{volume}{471}
  (\bibinfo{year}{2015}) \bibinfo{pages}{20140928}.
\bibitem[{Barnes et~al.(2021)Barnes, Willey, Rosenberg, Medina, and
  Juhl}]{barnes2021initial}
\bibinfo{author}{C.~J. Barnes}, \bibinfo{author}{C.~L. Willey},
  \bibinfo{author}{K.~Rosenberg}, \bibinfo{author}{A.~Medina},
  \bibinfo{author}{A.~T. Juhl},
\newblock \bibinfo{title}{Initial computational investigation toward passive
  transition delay using a phononic subsurface},
\newblock in: \bibinfo{booktitle}{AIAA Scitech 2021 Forum}, p.
  \bibinfo{pages}{1454}.
\bibitem[{Bonhomme et~al.(2019)Bonhomme, Oudich, Djafari-Rouhani, Sarry,
  Pennec, Bonello, Beyssen, and Charette}]{bonhomme2019love}
\bibinfo{author}{J.~Bonhomme}, \bibinfo{author}{M.~Oudich},
  \bibinfo{author}{B.~Djafari-Rouhani}, \bibinfo{author}{F.~Sarry},
  \bibinfo{author}{Y.~Pennec}, \bibinfo{author}{B.~Bonello},
  \bibinfo{author}{D.~Beyssen}, \bibinfo{author}{P.~Charette},
\newblock \bibinfo{title}{Love waves dispersion by phononic pillars for
  nano-particle mass sensing},
\newblock \bibinfo{journal}{Applied Physics Letters} \bibinfo{volume}{114}
  (\bibinfo{year}{2019}) \bibinfo{pages}{013501}.
\bibitem[{Dudley and Dudley(2012)}]{hensley_contfrac}
\bibinfo{author}{U.~Dudley}, \bibinfo{author}{U.~Dudley},
  \bibinfo{title}{{Continued Fractions}}, volume~\bibinfo{volume}{20},
  \bibinfo{publisher}{World Scientific}, \bibinfo{year}{2012}.
\bibitem[{Al~Ba'ba'a et~al.(2017)Al~Ba'ba'a, Nouh, and Singh}]{al2017pole}
\bibinfo{author}{H.~Al~Ba'ba'a}, \bibinfo{author}{M.~Nouh},
  \bibinfo{author}{T.~Singh},
\newblock \bibinfo{title}{Pole distribution in finite phononic crystals:
  Understanding bragg-effects through closed-form system dynamics},
\newblock \bibinfo{journal}{The Journal of the Acoustical Society of America}
  \bibinfo{volume}{142} (\bibinfo{year}{2017}) \bibinfo{pages}{1399--1412}.
\bibitem[{Yueh(2005)}]{yueh2005eigenvalues}
\bibinfo{author}{W.~C. Yueh},
\newblock \bibinfo{title}{{Eigenvalues of several tridiagonal matrices}},
\newblock \bibinfo{journal}{Applied Mathematics E - Notes} \bibinfo{volume}{5}
  (\bibinfo{year}{2005}) \bibinfo{pages}{66--74}.
\bibitem[{da~Fonseca(2007)}]{da2007eigenvalues}
\bibinfo{author}{C.~M. da~Fonseca},
\newblock \bibinfo{title}{{On the eigenvalues of some tridiagonal matrices}},
\newblock \bibinfo{journal}{Journal of Computational and Applied Mathematics}
  \bibinfo{volume}{200} (\bibinfo{year}{2007}) \bibinfo{pages}{283--286}.
\bibitem[{Al~Ba'ba'a et~al.(2017)Al~Ba'ba'a, Nouh, and Singh}]{al2017formation}
\bibinfo{author}{H.~Al~Ba'ba'a}, \bibinfo{author}{M.~Nouh},
  \bibinfo{author}{T.~Singh},
\newblock \bibinfo{title}{Formation of local resonance band gaps in finite
  acoustic metamaterials: A closed-form transfer function model},
\newblock \bibinfo{journal}{Journal of Sound and Vibration}
  \bibinfo{volume}{410} (\bibinfo{year}{2017}) \bibinfo{pages}{429--446}.
\bibitem[{Al~Ba'ba'a et~al.(2018)Al~Ba'ba'a, Depauw, Singh, and
  Nouh}]{al2018dispersion}
\bibinfo{author}{H.~Al~Ba'ba'a}, \bibinfo{author}{D.~Depauw},
  \bibinfo{author}{T.~Singh}, \bibinfo{author}{M.~Nouh},
\newblock \bibinfo{title}{{Dispersion transitions and pole-zero characteristics
  of finite inertially amplified acoustic metamaterials}},
\newblock \bibinfo{journal}{Journal of Applied Physics} \bibinfo{volume}{123}
  (\bibinfo{year}{2018}) \bibinfo{pages}{105106}.
\bibitem[{Sugino et~al.(2016)Sugino, Leadenham, Ruzzene, and
  Erturk}]{sugino2016mechanism}
\bibinfo{author}{C.~Sugino}, \bibinfo{author}{S.~Leadenham},
  \bibinfo{author}{M.~Ruzzene}, \bibinfo{author}{A.~Erturk},
\newblock \bibinfo{title}{{On the mechanism of bandgap formation in locally
  resonant finite elastic metamaterials}},
\newblock \bibinfo{journal}{Journal of Applied Physics} \bibinfo{volume}{120}
  (\bibinfo{year}{2016}) \bibinfo{pages}{134501}.
\bibitem[{Bastawrous and Hussein(2022)}]{bastawrous2022closed}
\bibinfo{author}{M.~V. Bastawrous}, \bibinfo{author}{M.~I. Hussein},
\newblock \bibinfo{title}{Closed-form existence conditions for bandgap
  resonances in a finite periodic chain under general boundary conditions},
\newblock \bibinfo{journal}{The Journal of the Acoustical Society of America}
  \bibinfo{volume}{151} (\bibinfo{year}{2022}) \bibinfo{pages}{286--298}.
\bibitem[{Targoff(1947)}]{targoff1947associated}
\bibinfo{author}{W.~P. Targoff},
\newblock \bibinfo{title}{The associated matrices of bending and coupled
  bending-torsion vibrations},
\newblock \bibinfo{journal}{Journal of the Aeronautical Sciences}
  \bibinfo{volume}{14} (\bibinfo{year}{1947}) \bibinfo{pages}{579--582}.
\bibitem[{Maradudin and Weiss(1958)}]{maradudin1958vibrations}
\bibinfo{author}{A.~Maradudin}, \bibinfo{author}{G.~H. Weiss},
\newblock \bibinfo{title}{On the vibrations of a generalized diatomic lattice},
\newblock \bibinfo{journal}{The Journal of Chemical Physics}
  \bibinfo{volume}{29} (\bibinfo{year}{1958}) \bibinfo{pages}{631--634}.
\bibitem[{Hori and Asahi(1957)}]{hori1957vibration}
\bibinfo{author}{J.-i. Hori}, \bibinfo{author}{T.~Asahi},
\newblock \bibinfo{title}{On the vibration of disordered linear lattice},
\newblock \bibinfo{journal}{Progress of Theoretical Physics}
  \bibinfo{volume}{17} (\bibinfo{year}{1957}) \bibinfo{pages}{523--542}.
\bibitem[{Maradudin and Weiss(1958)}]{maradudin1958disordered}
\bibinfo{author}{A.~Maradudin}, \bibinfo{author}{G.~H. Weiss},
\newblock \bibinfo{title}{The disordered lattice problem: A review},
\newblock \bibinfo{journal}{Journal of the Society for Industrial and Applied
  Mathematics} \bibinfo{volume}{6} (\bibinfo{year}{1958})
  \bibinfo{pages}{302--319}.
\bibitem[{Lin and McDaniel(1969)}]{lin1969dynamics}
\bibinfo{author}{Y.-K. Lin}, \bibinfo{author}{T.~McDaniel},
\newblock \bibinfo{title}{Dynamics of beam-type periodic structures}
  (\bibinfo{year}{1969}).
\bibitem[{Mead(1971)}]{mead1971vibration}
\bibinfo{author}{D.~J. Mead},
\newblock \bibinfo{title}{Vibration response and wave propagation in periodic
  structures}  (\bibinfo{year}{1971}).
\bibitem[{Xu et~al.(2012)Xu, Wu, and Guo}]{xu2012low}
\bibinfo{author}{Z.~Xu}, \bibinfo{author}{F.~Wu}, \bibinfo{author}{Z.~Guo},
\newblock \bibinfo{title}{Low frequency phononic band structures in
  two-dimensional arc-shaped phononic crystals},
\newblock \bibinfo{journal}{Physics Letters A} \bibinfo{volume}{376}
  (\bibinfo{year}{2012}) \bibinfo{pages}{2256--2263}.
\bibitem[{Hvatov and Sorokin(2015)}]{hvatov2015free}
\bibinfo{author}{A.~Hvatov}, \bibinfo{author}{S.~Sorokin},
\newblock \bibinfo{title}{{Free vibrations of finite periodic structures in
  pass- and stop-bands of the counterpart infinite waveguides}},
\newblock \bibinfo{journal}{Journal of Sound and Vibration}
  \bibinfo{volume}{347} (\bibinfo{year}{2015}) \bibinfo{pages}{200--217}.
\bibitem[{Carneiro~Jr et~al.(2021)Carneiro~Jr, Brennan, Gon{\c{c}}alves,
  Cleante, Bueno, and Santos}]{carneiro2021attenuation}
\bibinfo{author}{J.~Carneiro~Jr}, \bibinfo{author}{M.~Brennan},
  \bibinfo{author}{P.~Gon{\c{c}}alves}, \bibinfo{author}{V.~Cleante},
  \bibinfo{author}{D.~Bueno}, \bibinfo{author}{R.~Santos},
\newblock \bibinfo{title}{On the attenuation of vibration using a finite
  periodic array of rods comprised of either symmetric or asymmetric cells},
\newblock \bibinfo{journal}{Journal of Sound and Vibration}
  \bibinfo{volume}{511} (\bibinfo{year}{2021}) \bibinfo{pages}{116217}.
\bibitem[{Hussein et~al.(2006)Hussein, Hulbert, and
  Scott}]{hussein2006dispersive}
\bibinfo{author}{M.~I. Hussein}, \bibinfo{author}{G.~M. Hulbert},
  \bibinfo{author}{R.~A. Scott},
\newblock \bibinfo{title}{Dispersive elastodynamics of 1d banded materials and
  structures: analysis},
\newblock \bibinfo{journal}{Journal of sound and vibration}
  \bibinfo{volume}{289} (\bibinfo{year}{2006}) \bibinfo{pages}{779--806}.
\bibitem[{Hussein et~al.(2007)Hussein, Hulbert, and
  Scott}]{hussein2007dispersive}
\bibinfo{author}{M.~I. Hussein}, \bibinfo{author}{G.~M. Hulbert},
  \bibinfo{author}{R.~A. Scott},
\newblock \bibinfo{title}{Dispersive elastodynamics of 1d banded materials and
  structures: design},
\newblock \bibinfo{journal}{Journal of Sound and Vibration}
  \bibinfo{volume}{307} (\bibinfo{year}{2007}) \bibinfo{pages}{865--893}.
\bibitem[{Cheng et~al.(2018)Cheng, Shi, and Mo}]{cheng2018complex}
\bibinfo{author}{Z.~Cheng}, \bibinfo{author}{Z.~Shi}, \bibinfo{author}{Y.-L.
  Mo},
\newblock \bibinfo{title}{Complex dispersion relations and evanescent waves in
  periodic beams via the extended differential quadrature method},
\newblock \bibinfo{journal}{Composite Structures} \bibinfo{volume}{187}
  (\bibinfo{year}{2018}) \bibinfo{pages}{122--136}.
\bibitem[{Ruzzene and Baz(2000)}]{Ruzzene2000}
\bibinfo{author}{M.~Ruzzene}, \bibinfo{author}{A.~Baz},
\newblock \bibinfo{title}{{Control of wave propagation in periodic composite
  rods using shape memory inserts}},
\newblock \bibinfo{journal}{Journal of Vibration and Acoustics, Transactions of
  the ASME} \bibinfo{volume}{122} (\bibinfo{year}{2000})
  \bibinfo{pages}{151--159}.
\bibitem[{Petyt(2010)}]{petyt2010introduction}
\bibinfo{author}{M.~Petyt}, \bibinfo{title}{Introduction to finite element
  vibration analysis}, \bibinfo{publisher}{Cambridge university press},
  \bibinfo{year}{2010}.
\bibitem[{Curtain and Morris(2009)}]{curtain2009transfer}
\bibinfo{author}{R.~Curtain}, \bibinfo{author}{K.~Morris},
\newblock \bibinfo{title}{Transfer functions of distributed parameter systems:
  A tutorial},
\newblock \bibinfo{journal}{Automatica} \bibinfo{volume}{45}
  (\bibinfo{year}{2009}) \bibinfo{pages}{1101--1116}.
\bibitem[{Mead(1975)}]{mead1975wave}
\bibinfo{author}{D.~J. Mead},
\newblock \bibinfo{title}{{Wave propagation and natural modes in periodic
  systems: II. Multi-coupled systems, with and without damping}},
\newblock \bibinfo{journal}{Journal of Sound and Vibration}
  \bibinfo{volume}{40} (\bibinfo{year}{1975}) \bibinfo{pages}{19--39}.
\bibitem[{Djafari-Rouhani et~al.(1983)Djafari-Rouhani, Dobrzynski, Duparc,
  Camley, and Maradudin}]{djafari1983sagittal}
\bibinfo{author}{B.~Djafari-Rouhani}, \bibinfo{author}{L.~Dobrzynski},
  \bibinfo{author}{O.~H. Duparc}, \bibinfo{author}{R.~Camley},
  \bibinfo{author}{A.~Maradudin},
\newblock \bibinfo{title}{Sagittal elastic waves in infinite and semi-infinite
  superlattices},
\newblock \bibinfo{journal}{Physical Review B} \bibinfo{volume}{28}
  (\bibinfo{year}{1983}) \bibinfo{pages}{1711}.
\bibitem[{Shukla et~al.(2015)Shukla, Prasad, and Singh}]{shukla2015properties}
\bibinfo{author}{S.~Shukla}, \bibinfo{author}{S.~Prasad},
  \bibinfo{author}{V.~Singh},
\newblock \bibinfo{title}{Properties of surface modes in one dimensional plasma
  photonic crystals},
\newblock \bibinfo{journal}{Physics of Plasmas} \bibinfo{volume}{22}
  (\bibinfo{year}{2015}) \bibinfo{pages}{022122}.

\end{thebibliography}











\end{document}